\documentclass[aps,prb,reprint]{revtex4-2}

\usepackage{amsmath,amsfonts,amssymb,graphicx,braket,bbold,mathtools,array}
\usepackage{float}
\usepackage{multirow}
\usepackage[]{units}
\usepackage{hyperref}
\hypersetup{
    colorlinks=true,
    linkcolor=blue,
    filecolor=blue,      
    urlcolor=blue,
    citecolor=red
}

\usepackage[table]{xcolor}

\usepackage[T1]{fontenc}
\usepackage{tikz}
\usetikzlibrary{calc,patterns,decorations.pathmorphing,decorations.markings, shapes, positioning, shapes.geometric}

\newcommand*{\figref}[2][]{%
  \hyperref[{fig:#2}]{%
   \ref*{fig:#2}%
    \ifx\\#1\\%
    \else #1%
    \fi
  }%
}

\begin{document}
\title{All-electrical control of hole singlet-triplet spin qubits at low leakage points}
\author{Philipp M. Mutter}
\email{philipp.mutter@uni-konstanz.de}
\author{Guido Burkard}
\email{guido.burkard@uni-konstanz.de}
\affiliation{Department of Physics, University of Konstanz, D-78457 Konstanz, Germany}

\begin{abstract}
We study the effect of the spin-orbit interaction on heavy holes confined in a double quantum dot in the presence of a magnetic field of arbitrary direction. Rich physics arise as the two hole states of different spin are not only coupled by the spin-orbit interaction but additionally by the effect of site-dependent anisotropic $g$ tensors. It is demonstrated that these effects may counteract in such a way as to cancel the coupling at certain detunings and tilting angles of the magnetic field. This feature may be used in singlet-triplet qubits to avoid leakage errors and implement an electrical spin-orbit switch, suggesting the possibility of task-tailored  two-axes control. Additionally, we investigate systems with a strong spin-orbit interaction at weak magnetic fields. By exact diagonalization of the dominant Hamiltonian we find that the magnetic field may be chosen such that the qubit ground state is mixed only within the logical subspace for realistic system parameters, hence reducing leakage errors and providing reliable control over the qubit.

\end{abstract}
\maketitle
\section{Introduction}
Lately, heavy holes (HHs) confined in quantum dots (QDs) have gained ground in the race for a first scalable platform for quantum computation~\cite{Hendrickx2020, Hendrickx2020b,Wang2020arXiv, Lawrie2020,van_Riggelen2021, Hendrickx2021four,Wang2021}. Different implementations such as HHs in single QDs~\cite{Mutter2020cavitycontrol} and flopping mode qubits~\cite{Mutter2021natural} have been shown to allow for fast one and two qubit logic, externally controllable without the need for experimentally challenging components required in electronic systems such as oscillating magnetic fields or magnetic field gradients at the nano-scale~\cite{Benito2017, Mi2018, Benito2019a, Benito2019b, Croot2020}. 

\begin{figure}
    \includegraphics[scale=0.3]{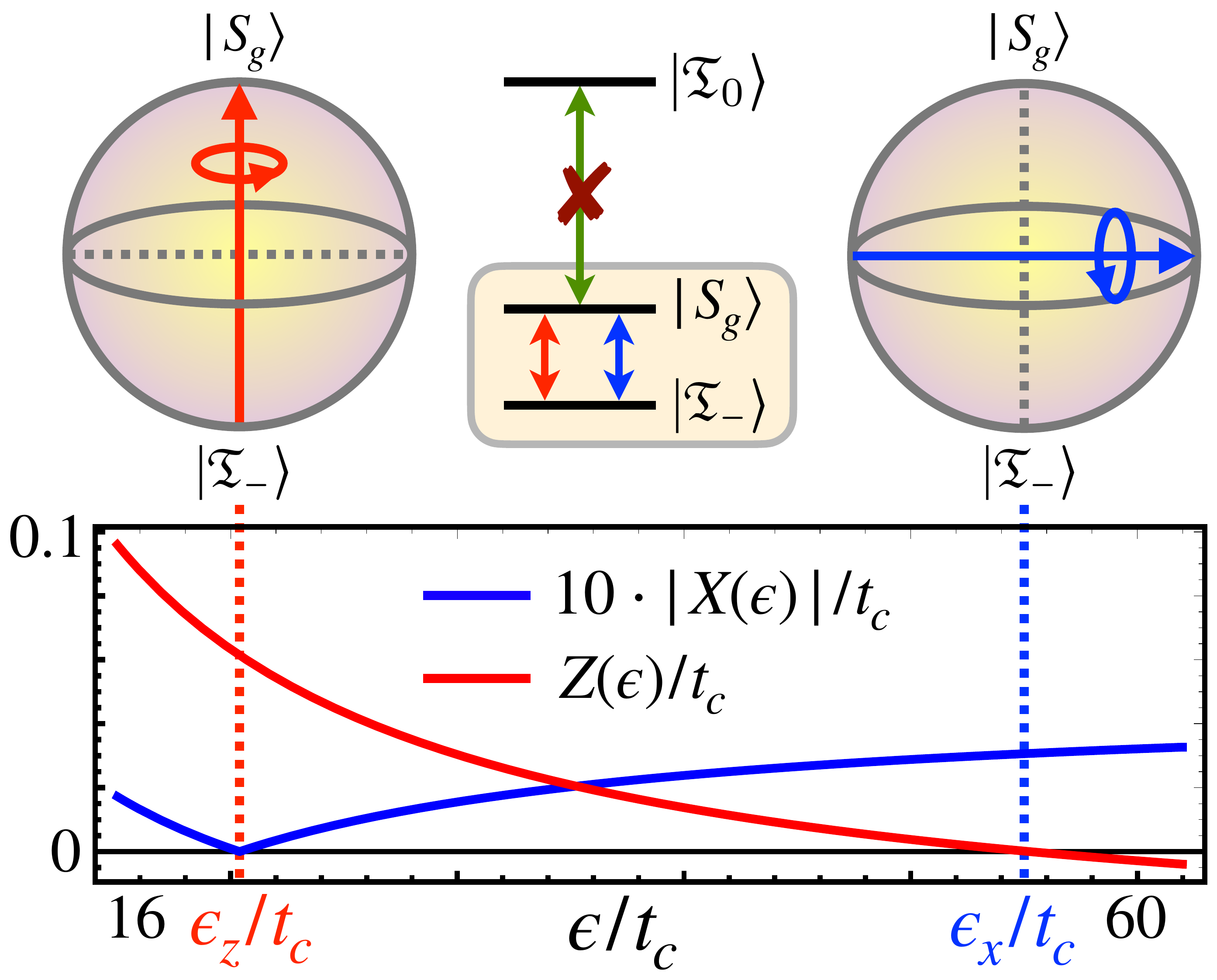}
    \caption{Electrically tunable two-axes control of a hole singlet-triplet qubit with logical states $\vert S_g \rangle$ and $\vert \mathfrak{T}_- \rangle$. By varying the double-dot detuning $\epsilon$ one may set either the coupling $X$ or the energy separation $Z$ to zero (bottom). The Hamiltonian then generates rotations around the $z$ and $x$ axes of the Bloch sphere, respectively, and the magnetic field direction can be fixed at an optimal point to reduce leakage errors (top). All quantities are displayed in units of the tunnel element $t_c \sim 10$~GHz, and their explicit forms in terms of the model parameters are derived in Sec.~\ref{sec:spin-orbit_switch}.}
    \label{fig:two_axes_control}
\end{figure}

Complete control of yet another promising qubit type, the singlet-triplet qubit~\cite{Levy2002, Petta2005}, often relies on magnetic field gradients~\cite{Foletti2009, Wu2014, Nichol2017} which require considerable effort for their experimental realization, or random nuclear fields~\cite{Koppens2005, Maune2012} which are hard to control. It was demonstrated that a singlet-triplet qubit can be realized if the double QD (DQD) system possesses site-dependent $g$ tensors~\cite{Jock2018}, which have been observed for holes in the group IV material germanium (Ge)~\cite{Hofmann2019arXiv}. Moreover, recent experiments found that a HH singlet-triplet qubit in planar Ge may be operated fast and coherently at magnetic fields below $10$~mT~\cite{Jirovec2021}. These achievements pave the way towards coupling Ge qubits to conventional superconductors such as aluminium in the context of super-semi hybrid circuit quantum electrodynamics \cite{Burkard2020}. While complete control was demonstrated with an applied out-of-plane magnetic field alone, it is desirable to fully understand the dependence on the direction of the magnetic field in such systems, e.g., to further increase qubit manipulation speed and coherence, as well as providing a reliable way to initialize and read out the system.
In this paper, we report the existence of special magnetic field directions in Ge DQDs where the leakage out of the hole-spin qubit subspace is suppressed and where all-electrical two-axes qubit control is possible (Fig.~\ref{fig:two_axes_control}).

There are three characteristic properties of HH systems, in particular in the semiconductor Ge: (i) Due to the p-symmetry of valence band orbitals, the contact hyperfine interaction of the hole spins with the nuclear spin bath is weak, thus reducing decoherence~\cite{Fischer2009,Fischer2010,Maier2012}. Additionally, since the nuclei in Ge predominantly have spin zero, it is possible to further reduce hyperfine interactions by isotopic purification~\cite{Sigillito2015}. (ii) There is a strong spin-orbit interaction (SOI) allowing for electrical spin control~\cite{Scappucci2020review}. In planar Ge, the SOI is of (cubic) Rashba type since the crystal possesses inversion symmetry~\cite{Bulaev2005, Bulaev2005b, Bulaev2007}. (iii) The $g$ tensors are highly anisotropic and can be  site-dependent~\cite{Hofmann2019arXiv,Jirovec2021}, yielding additional coupling terms and thus a higher degree of control. Here, we allow for a magnetic field of arbitrary direction and show that the combined effects of the SOI and anisotropic site-dependent $g$ tensors amount to a spin-orbit switch in a qubit that is at the same time protected from leakage errors. This is achieved in two steps: First, one fixes the magnetic field at a specific orientation depending on the system parameters which we specify, resulting in the reduction and in some cases even the elimination of leakage to states outside the logical subspace. With the magnetic field in place, the coupling between the qubit states can then be switched on and off electrically, hence offering optimal working conditions for different quantum computing tasks such as spin manipulation (SOI turned on) and read-out (SOI turned off).

After introducing the model in Sec.~\ref{sec:model}, we explicitly determine the optimal working points and derive effective singlet-triplet qubit Hamiltonians in Sec.~\ref{sec:spin-orbit_switch}. When the system is operated at an optimal point, we find $H_{\text{eff}} = -Z(\epsilon) \vert S_g \rangle \langle S_g \vert + \left( X(\epsilon) \vert S_g \rangle \langle \mathfrak{T}_- \vert + \text{H.c.} \right) $, where the ground state singlet $\vert S_g \rangle$ and the triplet $\vert \mathfrak{T}_- \rangle$ are the logical qubit states. The quantities $Z(\epsilon)$ and $X(\epsilon)$ can be switched on and off by the adjusting the double-dot energy detuning $\epsilon$ (Fig.~\ref{fig:two_axes_control}), hence offering all-electrical two-axes control over the qubit at a low leakage point. Since the axes are orthogonal, any one-qubit gate can be realized using three rotations~\cite{Barenco1995}, avoiding the more complicated control sequences required for non-orthogonal axes~\cite{Hanson2007}. We proceed to consider the case of a strong SOI in Sec.~\ref{sec:SO_switch_strong_SOI} and find that many of the desirable properties described in Sec.~\ref{sec:spin-orbit_switch} still hold in this case. Finally, Sec.~\ref{sec:conclusion} provides a conclusion.

\section{Model Hamiltonian}
\label{sec:model}

\begin{figure}
		\includegraphics[scale  = 0.195 ]{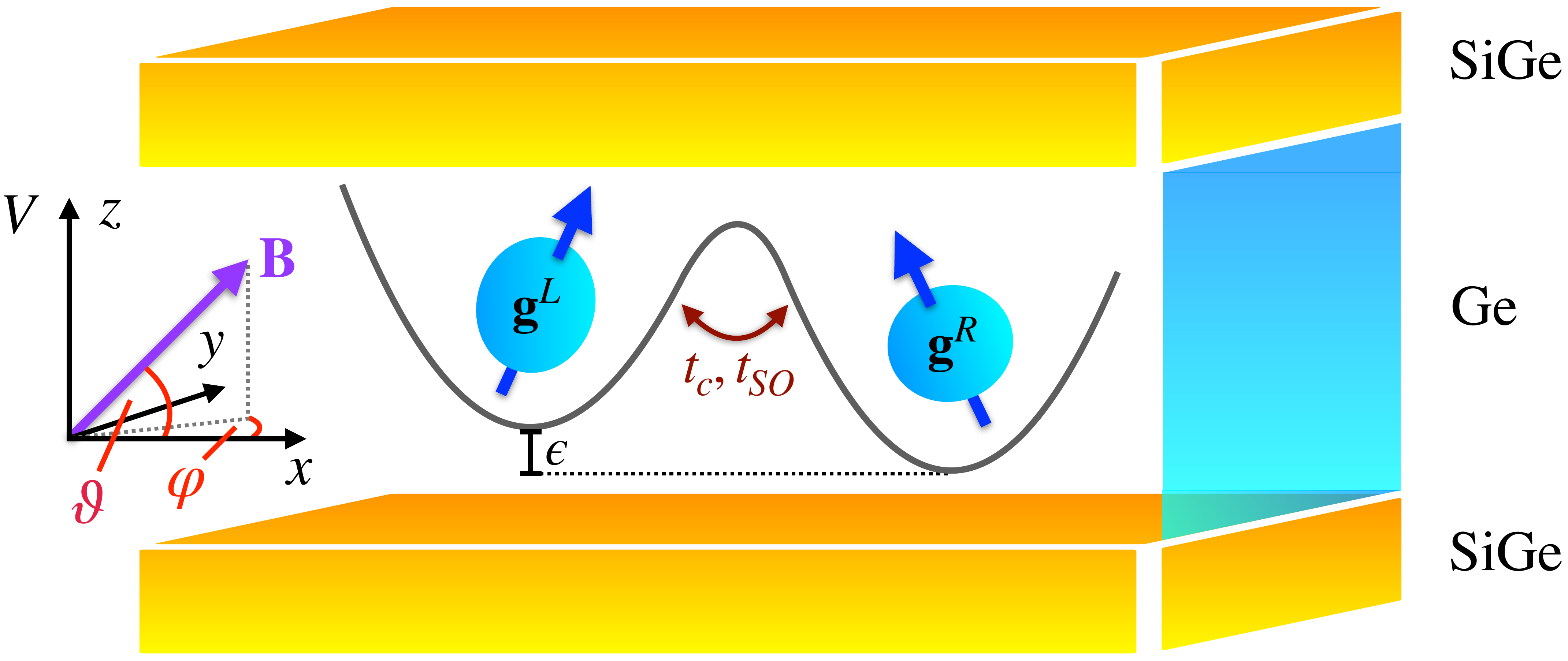}
		\caption{A planar DQD system. Holes confined to the middle layer of a Ge-SiGe heterostructure are subjected to a double quantum well potential, thus forming a tunnel coupled planar DQD. The two hole spin states are affected by the SOI which induces spin-flip tunneling processes and dot-dependent $g$ tensors. We apply a magnetic field $\mathbf{B}$ with out-of-plane tilting angle $\vartheta$ and in-plane tilting angle $\varphi$ as measured from the DQD axis  $x$.}
		\label{fig:system_schematic}
	\end{figure}

We consider a tunnel-coupled DQD with single particle tunneling matrix element $t_c$ in a regime where the (0,2) singlet is far detuned. A typcial architecture realizing the system as a planar DQD in a Ge-GeSi heterostructure is shown in Fig.~\ref{fig:system_schematic}. Allowing for spin-flip tunneling processes induced by the SOI and an external magnetic field of arbitrary direction, the Hamiltonian reads~\cite{Jouravlev2006, Danon2009},
	\begin{align}
	\label{eq:model_Hamiltonian}
	\begin{split}
		& H =  H_0 + H_{\text{SO}} + H_Z, \\
		& H_0 = \epsilon \vert S_{20} \rangle \langle S_{20} \vert + \sqrt{2} t_c \left( \vert S \rangle \langle S_{20}  \vert + \vert S_{20} \rangle \langle S  \vert \right), \\
		& H_{\text{SO}} = i \sqrt{2} t_z \vert T_0 \rangle \langle S_{20} \vert - \sum_{\pm} (t_y \pm i  t_x) \vert T_{\pm} \rangle \langle S_{20} \vert + \text{H.c.},
	\end{split}
	\end{align}
where $\epsilon \geqslant 0$ is the detuning relative to the (1,1)-(2,0) singlet crossing, and $\mathbf{t}_{\text{SO}} = (t_x,t_y,t_z)$ is the spin-orbit vector of the system. Note that $\mathbf{t}_{\text{SO}}$ is scaled by a factor $1/ \sqrt{2}$ compared to Ref.~\cite{Danon2009} to unburden the notation and lighten the expressions later on. Information about the spin-orbit vector may be obtained, e.g., via magneto-transport measurements~\cite{Mutter2020, MutterPSB2021}. Higher order contributions due to the SOI are possible but suppressed when the orbital spacing induced by the confinement potentials is the largest energy scale in the system~\cite{Golovach2004}. Finally, the Zeeman Hamiltonian $H_Z$ contains anisotropic site-dependent $g$ tensors which lead to additional coupling terms between singlets and triplets. In general, our analysis applies whenever the two $g$ tensors are diagonal in a common eigenbasis. In the following we take the $g$ tensor in each dot to be diagonal in the basis defined by the cubic crystal and degenerate in the $x$-$y$-plane, $\mathbf{g} = \text{diag}(g_x,g_x,g_z)$. In all the plots in this paper, the $g$ tensor components are taken to have values typical for HHs in Ge, $g_x^L = 0.2$, $g_x^R = 0,3$, $g_z^L = 6.5$ and $g_z^R = 4.5$, where $L/R$ labels the left/right dot. Parametrizing the magnetic field in spherical coordinates with the $x$-$y$-plane as the equatorial plane and the azimuthal angle $\varphi$ measured from the DQD axis~$x$, $\mathbf{B} = B (\cos \vartheta \cos \varphi, \cos \vartheta \sin \varphi, \sin \vartheta)$, the Zeeman Hamiltonian has the form
	\begin{align}
	\label{eq:HZinSTbasis}
		\begin{split}
			&H_{\text{Z}} = \frac{B \sin \vartheta}{2} \left[ \sum_{\pm} \pm g_z^+  \vert T_{\pm} \rangle \langle T_{\pm} \vert  +  g_z^- \left( \vert S \rangle \langle T_0 \vert +\text{H.c.}\right) \right] \\
			&+  \frac{B \cos \vartheta}{2 \sqrt{2}}  \bigg[  \sum_{\pm} e^{\pm  i \varphi} \left( g_x^+  \vert T_0 \rangle \langle T_{\pm} \vert  \mp   g_x^-   \vert S \rangle \langle T_{\pm} \vert \right) + \text{H.c.} \bigg],
		\end{split}
	\end{align}
where $g_x^{\pm} = g_x^L \pm g_x^R$ and $g_z^{\pm} = g_z^L \pm g_z^R$ are the sums and differences of the $g$ tensor components in the left and right dot.

The proposed model is rather general and may in principle be applied to a wide range of materials. However, the presence of site-dependent and highly anisotropic $g$ tensors in combination with a strong SOI is typical for HH states, e.g., in the semiconductor Ge. Moreover, the specialization of HH systems allows us to neglect interactions of the hole spins with the nuclear spin bath.
	
\subsection{Dominant basis}

Although the SOI and the site-dependence of the $g$ tensors are non negligible effects in HH systems, the largest energy scale is still expected to be given by the spin conserving tunnel coupling $t_c$ and the standard Zeeman terms featuring the sum of $g$ factors in the dots. The case where the spin-flip tunneling terms induced by the SOI are of the same order of magnitude as $t_c$ is discussed in Sec.~\ref{sec:SO_switch_strong_SOI}. For now we work in the regime where the dominant part of the Hamiltonian is $H$ as given in Eq.~\eqref{eq:model_Hamiltonian} but for vanishing SOI ($\mathbf{t}_{\text{SO}} =0 $) and equal $g$ tensors ($\mathbf{g}^L = \mathbf{g}^R$). It is diagonal in the states, 
	\begin{align}
	\label{eq:hybridized_basis}
	\begin{split}
		& \vert S_e \rangle = \sin \frac{\Omega}{2}   \vert S \rangle+ \cos \frac{\Omega}{2}  \vert S_{20} \rangle , \\
		& \vert S_g \rangle = \cos \frac{\Omega}{2}    \vert S \rangle - \sin \frac{\Omega}{2}  \vert S_{20} \rangle, \\
			& \vert \mathfrak{T}_0 \rangle = f_z \vert T_0 \rangle -  \frac{f_x}{\sqrt{2}}  \sum_{\nu = \pm} \nu e^{- \nu i \varphi} \vert T_{\nu} \rangle, \\
			& \vert \mathfrak{T}_{\pm} \rangle = \sum_{\nu = \pm} \frac{1 \pm \nu f_z}{2} e^{- \nu i \varphi}  \vert T_{\nu} \rangle  \pm   \frac{f_x}{\sqrt{2}}   \vert T_0 \rangle,		
	\end{split}
	\end{align}
where we introduce the orbital hybridization angle $\Omega = \arctan \left(2\sqrt{2}t_c/ \epsilon \right)$ for the hybridized singlets and the dimensionless functions $ f_x = g_x^+ \cos \vartheta/G_+$,  $ f_z = g_z^+ \sin \vartheta/G_+$ with the effective sum of $g$ factors,
    \begin{align}
        G_+ = \sqrt{  ( g_x^+ \cos \vartheta)^2 +  ( g_z^+ \sin \vartheta)^2},
    \end{align}
for the mixed triplet states. Transforming $H$ into the basis $\lbrace \vert S_e \rangle, \vert S_g \rangle,\vert \mathfrak{T}_0 \rangle, \vert \mathfrak{T}_+ \rangle, \vert \mathfrak{T}_- \rangle \rbrace$ yields a Hamiltonian in which the ground state singlet $\vert S_g \rangle$ is separated from the triplets by the exchange energy $J = \sqrt{\epsilon^2/4 + 2 t_c^2} - \epsilon/2$, and which features the relevant singlet triplet mixing terms,

	\begin{align}
	\label{eq:H_hybridized_basis}
		 \tilde{H}  =& \begin{pmatrix}
			\epsilon + J & 0 & i \sqrt{2} M_0^+ & M_+^+ & M_-^+ \\
			0 & -J &- i \sqrt{2} M_0^- & -M_+^-& - M_-^- \\
			-i \sqrt{2} M_0^+ & i \sqrt{2} M_0^- & 0 & 0 & 0 \\
			M_-^+ & - M_-^- & 0 & 0 & 0 \\
			M_+^+ & -M_+^- & 0 & 0 & 0 \\
		\end{pmatrix}\nonumber	\\ 	
		& + \frac{ B}{2} 
		\begin{pmatrix}
			0 & 0 & G_-^- &G_a^- & -G_a^-  \\
			0 & 0 & G_-^+ &G_a^+ & -G_a^+ \\
			G_-^- & G_-^+  & 0 & 0 & 0 \\
			G_a^- & G_a^+  & 0 & G_+ & 0 \\
			-G_a^- & -G_a^+ & 0 & 0 & -G_+ \\
		\end{pmatrix},
		\end{align}
		
where $X^{\pm} = X \sqrt{(1 \pm \cos \Omega)/2}$ for upper indices $\pm$. The matrix elements due to the SOI are given by
	\begin{align}
	\label{eq:matrix_elements_SOI}
	\begin{split}
		& M_0 = t_z f_z + f_x \left(t_x \cos \varphi + t_y \sin \varphi \right), \\
		& M_{\pm} = \pm i t_z f_x  +  \frac{1}{2i}  \sum_{\nu = \pm 1}  (\nu t_x -  it_y)(1 \pm \nu f_z)e^{i \nu \varphi} ,
	\end{split}
	\end{align}
	and the Hermiticity of the Hamiltonian is warranted by
$M_{\pm}^* = M_{\mp}$, while the effective two-particle $g$ factors read
	\begin{align}
	\label{eq:effective_g_factors}
	\begin{split}
		& G_- = \frac{1}{G_+} \left( g_x^-g_x^+ \cos^2 \vartheta + g_z^-g_z^+ \sin^2 \vartheta \right)  , \\ 	
		&G_a = \frac{1}{2\sqrt{2}G_+} \left(  g_x^+ g_z^- - g_x^- g_z^+ \right) \sin 2 \vartheta .
	\end{split}
	\end{align}
The term $G_a$ arises due to the anisotropy of the $g$ tensors and is only present when the magnetic field has non-zero in- and out-of-plane components, $\vartheta \in (0, \pi/2)$, i.e., when the field is not aligned with one of the principal axes of the $g$ tensor. Put differently, the term is a consequence of non-parallel effective magnetic fields $\mathfrak{B}^d_i = \sum_j g^d_{ij} B_j$ in the dots $d \in \lbrace L,R \rbrace$, which enclose an angle
	\begin{align}
		\theta^d = \arctan \left( \frac{g_z^d}{g_x^d} \tan \vartheta \right)
	\end{align}		
with the equatorial plane. Due to the inverse effect of increased HH light hole (LH) mixing on the $g$ tensor components, i.e., a reduction of the out-of-plane $g$ factor and an enhancement of the in-plane $g$ factor, one expects $g_x^L > g_x^R$ for $g_z^L < g_z^R$ and vice versa. Hence, the anisotropy term $G_a$ can be quite sizeable when the $g$ tensors are different. For isotropic $g$ tensors one has $G_+ = g^+$, $G_- = g^-$ and $G_a = 0$.	
	
It is worth pointing out that the energies of the dominant eigenstates and the couplings between these states obtained by diagonalization of the dominant part of the Hamiltonian and the transformation of the non-dominant part into this basis agree with those obtained by naively choosing the global quantization axis along the sum of effective magnetic fields $\boldsymbol{\mathfrak{B}}^L +  \boldsymbol{\mathfrak{B}}^R$ as detailed in Appendix~\ref{appx:quantization_axis}. This result is non-trivial as the quantization axis is fixed to be out-of-plane in two dimensional holes systems as a direct consequence of the spin-momentum locking in the Luttinger-Kohn Hamiltonian and the position-momentum uncertainty relation~\cite{Scappucci2020review}.

\subsection{Symmetries of the system}
\label{sec:symmetries_of_the_system}

The system possesses two symmetries which are not obvious from the Hamiltonian~\eqref{eq:H_hybridized_basis}: time-reversal symmetry in combination with the inversion of the magnetic field and symmetry under the exchange of the two dots.
 
Firstly we note that under the combined inversion of time, $t \rightarrow - t$, and the magnetic field, $\mathbf{B} \rightarrow - \mathbf{B}$ (i.e., $\vartheta \rightarrow - \vartheta, \; \varphi \rightarrow \varphi + \pi \mod 2 \pi$), one has  $\vert \mathfrak{T}_0 \rangle \rightarrow  - \vert \mathfrak{T}_0 \rangle$ and $ \vert \mathfrak{T}_{\pm} \rangle \rightarrow  \vert \mathfrak{T}_{\pm} \rangle$. Since furthermore $i M_0 \rightarrow i M_0$ and $M_{\pm}  \rightarrow -M_{\pm} $, the total Hamiltonian~\eqref{eq:H_hybridized_basis} is invariant, in particular for $B = 0$, where it is time-reversal invariant.

\begin{figure}
		\includegraphics[scale=0.195]{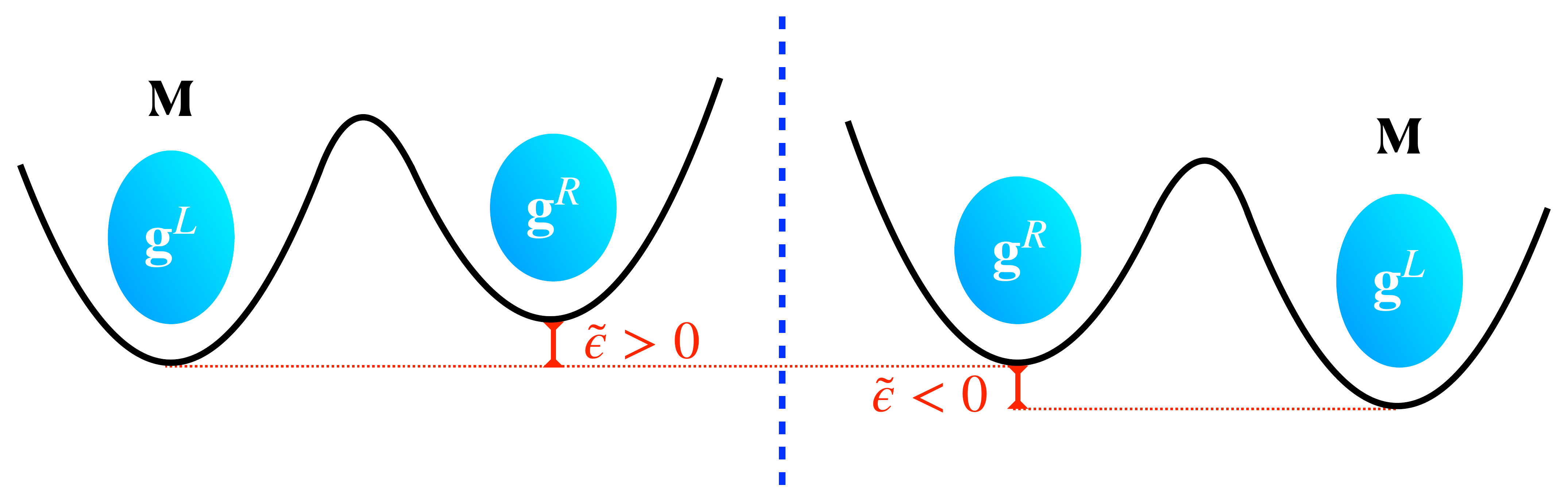}
		\caption{Symmetry transformation of the system that exchanges the left and right dot. The system is invariant when the left and right $g$ tensors are exchanged, the inter-dot detuning $\tilde{\epsilon}$ is inverted, and $\vert S_{02} \rangle$ instead of $\vert S_{20} \rangle$ is used as the measurement point $\mathbf{M}$. Up to a constant and hence physically irrelevant energy, the DQD setup can be regarded as being mirrored about the dashed blue line.}
		\label{fig:symmetry_L_R}
	\end{figure}

Secondly, we point out a symmetry under the exchange of the two dots. There is a peculiarity arising due to site-dependent anisotropic $g$ tensors. When exchanging the $g$ tensors, $\boldsymbol{\hat{g}}^L \leftrightarrow \boldsymbol{\hat{g}}^R$, one finds $G_- \rightarrow -G_-$ and $G_a \rightarrow -G_a$ which amounts to a relative minus in the coupling terms between the singlets and the triplets. As a consequence, the different $g$ tensors distinguish the left and right dot in a way that can be measured. To understand this observation, one must work in the enlarged six-dimensional Hilbert space, now containing both singlets with double dot occupancy, $\vert  S_{02} \rangle $ and $\vert  S_{20} \rangle$, with energies $U + \tilde{\epsilon}$ and $U - \tilde{\epsilon}$, respectively, $U$ being the charging energy and $\tilde{\epsilon}$ the energy detuning between the dots in the one particle picture. Note that the spin conserving tunnel matrix element has the same sign for tunneling events from $\vert S \rangle$ to both $\vert  S_{02} \rangle $ and $\vert  S_{20} \rangle $, while the spin-orbit matrix elements satisfy $\langle   S_{02}   \vert H_{\text{SO}} \vert T_{\nu} \rangle = - \langle   S_{20}   \vert H_{\text{SO}} \vert T_{\nu} \rangle$ for $\nu \in \lbrace 0, \pm \rbrace$. Consequently, the Hamiltonian is invariant with respect to exchanging the $g$ tensors if we furthermore invert the inter-dot detuning $\tilde{\epsilon}$ and relabel the states  $\vert S_{20} \rangle \leftrightarrow \vert S_{02} \rangle$~\footnote{In addition, one must rescale the triplet states by minus one. However, since the physical states are only representatives of an entire ray of states in a Hilbert space, this minus only amounts to an irrelevant phase.}. The combination of these transformations amounts to a complete reversal of the system architecture (Fig.~\ref{fig:symmetry_L_R}). Experimentally, the change of the effective two-particle $g$ factors when exchanging the one-particle $g$ tensors reflects the fact that one arbitrarily chooses one site as the (doubly occupied) measurement point by choosing the sign of the inter dot detuning, e.g., $\vert S_{20} \rangle$ for $\tilde{\epsilon} > 0$ in this work.

\section{The spin-orbit switch}
\label{sec:spin-orbit_switch}
A spin-orbit switch is present in a system if the SOI can be turned on and off by changing external control parameters such as gate voltages or an applied magnetic field. When operating such a system one may utilize the SOI for one application, e.g., for spin qubit manipulation in the context of spintronics, while one may choose to turn it off for another, e.g., for the purpose of spin readout or for assessing the effect of (residual) nuclear spins in the system. Therefore, a spin-orbit switch is a highly desirable property in platforms used for quantum information processing. Recently, this functionality was reported for holes in a Ge/silicon core/shell nanowire~\cite{Froning2021}.

\begin{figure}
        \includegraphics[scale=0.27]{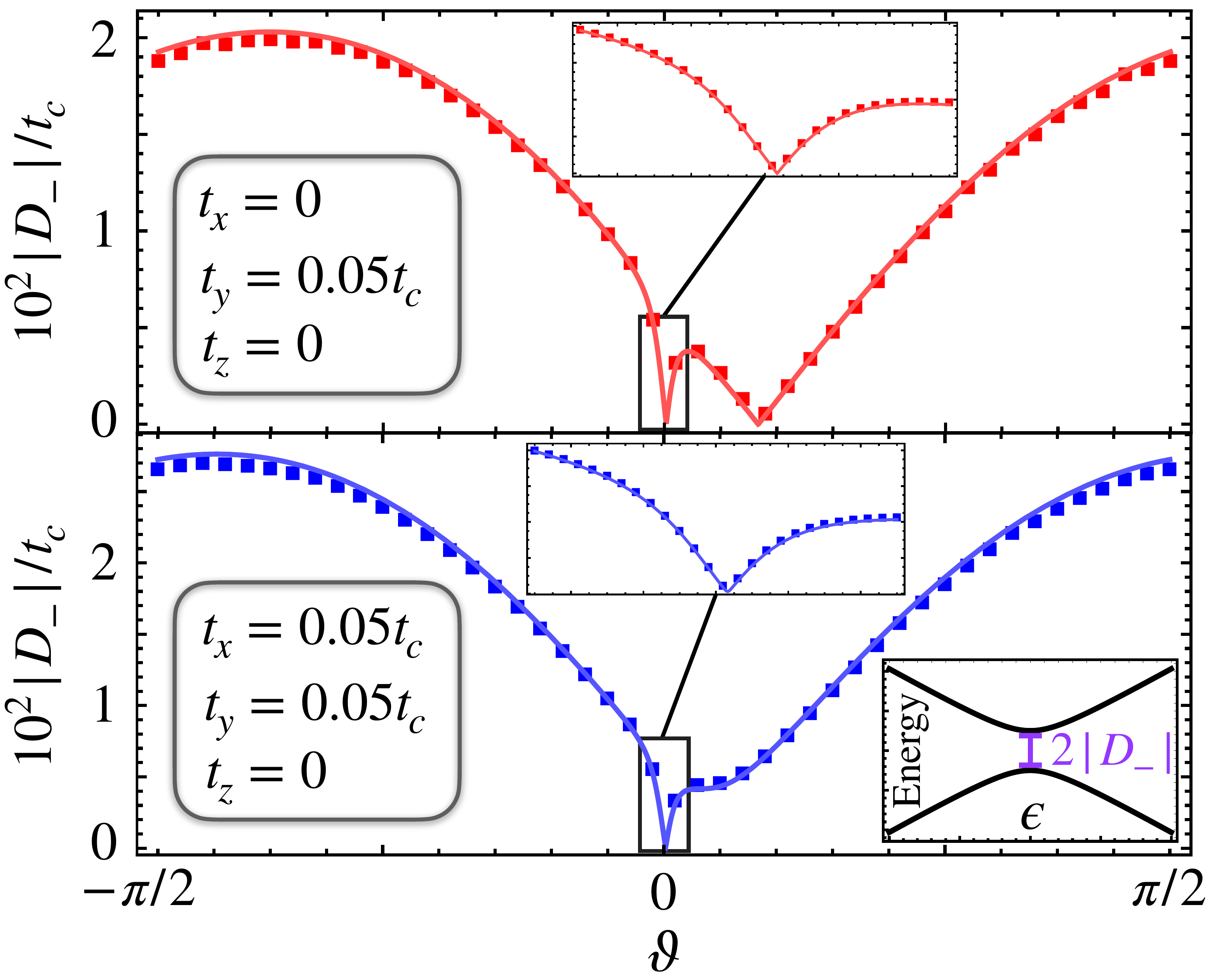}
        \caption{Validity of the model. We show a comparison between the analytical result for the coupling $\vert D_- \vert$ defined in Eq.~\eqref{eq:coupling_moderate_SOI} (solid lines, Eq.~\eqref{eq:dpm_effective_field}) and the exact result obtained by numerical diagonalization of the Hamiltonian~\eqref{eq:model_Hamiltonian} (squares) for two different spin-orbit vectors as indicated in the figure. We set $B=0.1 t_c$, $\varphi=0$ and the detuning is chosen such that the eigenstates of the dominant Hamiltonian $\vert S_g \rangle$ and $\vert \mathfrak{T}_- \rangle$ align in energy, $\epsilon = (8t_c^2 - (BG_+)^2)/2BG_+$. Numerically, the coupling is given by half the minimal energy difference between the two low-lying eigenstates of the full Hamiltonian as sketched in the lower right inset.}
        \label{fig:comparison_ana_num}
    \end{figure}

The logical singlet-triplet qubit space is spanned by the ground state singlet $\vert S_g \rangle$ and one of the hybridized triplet states $\vert \mathfrak{T}_{\nu} \rangle$. Hence, the relevant couplings are
	\begin{align}
	\label{eq:coupling_moderate_SOI}
		D_{\nu}  \equiv \langle S_g \vert \tilde{H} \vert \mathfrak{T}_{\nu} \rangle , \;  \nu \in \lbrace 0, \pm \rbrace.
	\end{align}
 In our approach the mixing terms are obtained by an exact basis transformation, and there has been no approximation made so far. Whether the actual physical system is accurately described by the expressions~\eqref{eq:coupling_moderate_SOI} depends on the validity of the choice of basis, the requirement being that the diagonal terms in the Hamiltonian matrix~\eqref{eq:H_hybridized_basis} dominate over the off-diagonal terms. In Fig.~\ref{fig:comparison_ana_num} we compare the analytical expression for $D_-$ to the exact result obtained by numerical diagonalization of the Hamiltonian~\eqref{eq:model_Hamiltonian}. We find excellent agreement, justifying our choice of basis even for relatively large out-of-plane $g$ factor differences ($g_z^- \sim 2$) and spin-flip tunneling terms ($t_{\text{SO}} \sim 0.05 t_c$). When the SOI becomes even larger, a different choice of basis can be more appropriate (Sec.~\ref{sec:SO_switch_strong_SOI}). Note that when a Schrieffer-Wolff transformation is performed to decouple the excited singlet $\vert S_e \rangle$ from the four dimensional low-energy space, the couplings $D_{\nu}$ are unchanged to leading order in the small quantities $\vert \mathbf{t}_{\text{SO}} \vert/\Delta E$, $B g_{x/z}^-/\Delta E$, where $\Delta E$ is the energy separation between the low-lying states and $\vert S_e \rangle$, and $\vert \mathbf{t}_{\text{SO}} \vert$, $B g_{x/z}^-$ are the mixing terms. Finally, we note that couplings of the form~\eqref{eq:coupling_moderate_SOI} may be picked up experimentally by measuring the singlet return probability in combination with Landau-Zener schemes~\cite{Petta2010}. Since all four low-energy states interact, it can be necessary to perform an additional Schrieffer-Wolff transformation to obtain an effective two-level system for which the standard Landau-Zener formalism can be applied.

As can be seen from Eq.~\eqref{eq:H_hybridized_basis}, there are two contributions to $D_{\nu}$ as defined in Eq.~\eqref{eq:coupling_moderate_SOI}: the SOI via the terms $M_0$ ($M_{\pm})$ and site-dependent $g$ tensors via $G_-$ ($G_a$) for $D_0$ ($D_{\pm}$). As we will see, these terms may counteract for certain magnetic field  angles and detunings such that $D_{\nu} = 0$, which effectively corresponds to a switched off SOI.

\subsection{Longitudinal coupling}
We first consider the coupling between the ground state singlet $\vert S_g \rangle$ and the unpolarized (longitudinal) triplet $\vert \mathfrak{T}_0 \rangle$. Introducing the sums and differences of effective magnetic fields for diagonal $g$ tensors in both dots, $\mathfrak{B}^{\pm}_j =  g_j^{\pm} B_j$, we may write
	\begin{align}
	\label{eq:d0_effective_field}
	\begin{split}
		& D_0 =     \mathbf{w}  \cdot   \boldsymbol{ \mathfrak{B}}^+_n,
	\end{split}
	\end{align}
where $\boldsymbol{ \mathfrak{B}}^+_n  = \boldsymbol{ \mathfrak{B}}^+/ \vert   \boldsymbol{ \mathfrak{B}}^+ \vert$ with $\vert   \boldsymbol{ \mathfrak{B}}^+ \vert = B G_+$ is the normalized version of $ \boldsymbol{ \mathfrak{B}}^+$ and
	\begin{align}
	\label{eq:w_expression}
		\mathbf{w} =  i \sqrt{2} \sin \left( \frac{\Omega}{2} \right) \mathbf{t}_{\text{SO}} + \frac{1}{2} \cos \left( \frac{\Omega}{2} \right)  \boldsymbol{ \mathfrak{B}}^-.
	\end{align}
Note that the scalar product in Eq.~\eqref{eq:d0_effective_field} is defined such that complex conjugation is performed in the first entry, $\mathbf{a} \cdot \mathbf{b} = \sum_j a^*_j b_j$. Since the spin-orbit vector is real, the vector $\mathbf{w}$ is non-zero for non-vanishing $ \mathbf{t}_{\text{SO}}$ and $\boldsymbol{ \mathfrak{B}}^-$, and we have $D_0 = 0$ only when the vectors $ \boldsymbol{ \mathfrak{B}}^+$ and $\mathbf{w}$ are orthogonal.

A necessary condition for $D_0 = 0$ is $G_-=0$ and hence
	\begin{align}
	\label{eq:zero_line_d0}
		\tan \vartheta = \pm \sqrt{-\frac{g_x^-g_x^+}{g_z^- g_z^+}}.
	\end{align}
Note that we expect $\text{sgn} (g_x^-) \neq \text{sgn} (g_z^-)$ as the out-of-plane $g$ factor decreases with stronger HH-LH mixing, while the in-plane $g$ factor increases. Hence, the radicand is positive and a real solution to Eq.~\eqref{eq:zero_line_d0} exists. One then finds the additional conditions for the azimuthal angle,
    \begin{align}
    \begin{split}
    \label{eq:zero_line_d0_phi}
        & \frac{t_x}{t_z}\cos \varphi + \frac{t_y}{t_z}\sin \varphi = \pm \sqrt{-\frac{g_x^-g_z^+}{g_z^- g_x^+}}, \quad t_z \neq 0, \\
        & \tan \varphi = - \frac{t_x}{t_y}, \quad t_z = 0.
    \end{split}
    \end{align}
Hence, the line of zero coupling is independent of the magnetic field strength and the detuning. A plot of the coupling for the case $t_z = 0$ is shown in Fig.~\figref[(a)]{D0_Dminus}.

\subsection{Transverse coupling}
We now turn to the couplings between the ground state singlet $\vert S_g \rangle$ and the polarized (transverse) triplets $\vert \mathfrak{T}_{\pm} \rangle$. A classification of magnetic degeneracy points in systems with site-dependent $g$ tensors and SOI was reported in Ref.~\cite{Gyorgy2020}. At these degeneracy points the ground state becomes degenerate in energy, corresponding to a vanishing coupling between the two lowest lying states, $D_- =0$ in our case. In this section, we investigate how such points can be controlled by the model parameters, and additionally investigate the zeros of the other transverse coupling $D_+$. Using the quantities introduced in Eqs.~\eqref{eq:d0_effective_field} and~\eqref{eq:w_expression}, one has
\begin{align}
	\label{eq:dpm_effective_field}
	\begin{split}
		& D_{\pm}  = \pm \frac{1}{\sqrt{2}}\left(\mathbf{w} \times  \boldsymbol{  \mathfrak{B}}^+_n \right) \cdot   \mathbf{v}_{\pm},
	\end{split}
	\end{align}
where $\mathbf{v}_{\pm} = (- \sin \varphi, \cos \varphi, \mp i/f_x)$. In contrast to the scalar product, the cross product is defined without complex conjugation, $(\mathbf{a} \times \mathbf{b})_k = \sum_{i,j} \epsilon_{ijk} a_i b_j$ with the totally anti-symmetric tensor $\epsilon_{ijk}$. Zero coupling is achieved for
	\begin{align}
	\label{eq:zero_line_Dpm}
		B G_a(\vartheta(\varphi))= \pm 2 \left(t_x \sin \varphi - t_y \cos \varphi  \right) \tan \frac{\Omega}{2},
	\end{align}
where $\pm$ corresponds to $D_{\pm}$ and the polar angle is expressed in terms of the azimuthal angle,
	\begin{align}
	\label{eq:theta_of_phi}
		 t_z \cot (\vartheta) =  \frac{g_z^+}{g_x^+} \left( t_x \cos \varphi + t_y \sin \varphi \right),
	\end{align}
For a given spin-orbit vector $\mathbf{t}_{\text{SO}}$ Eq.~\eqref{eq:zero_line_Dpm} can be used to determine the values of $B$, $\varphi$ and $\Omega$ for which the coupling vanishes. By substituting back into Eq.~\eqref{eq:theta_of_phi} one obtains the corresponding polar angle $\vartheta$. 

It is instructive to look at some special cases for the spin-orbit vector. If $\mathbf{t}_{\text{SO}} = (0,0,t_z)$, one has $\vartheta = \pi/2$ by Eq.~\eqref{eq:theta_of_phi} and hence Eq.~\eqref{eq:zero_line_Dpm} is satisfied independently of $B$, $\varphi$ and $\Omega$. On the other hand, if $\mathbf{t}_{\text{SO}} = (t_x,t_y,0)$, there are two configurations that lead to zero coupling: (i) If Eq.~\eqref{eq:theta_of_phi} is satisfied by setting $\vartheta = 0$, Eq.~\eqref{eq:zero_line_Dpm} requires $\tan \varphi = t_y/t_x $ since $\Omega \neq 0$ at finite detuning. (ii) Eq.~\eqref{eq:theta_of_phi} may also be satisfied by setting $\tan \varphi = - t_x/t_y$, remarkably the same azimuthal angle as required by Eq.~\eqref{eq:zero_line_d0_phi} for the longitudinal coupling $D_0$ to vanish.  Eq.~\eqref{eq:zero_line_Dpm} then yields
 	\begin{align}
	\label{eq:Dpm_ty}
		BG_a(\vartheta) = \mp 2t_{\text{SO}} \tan \frac{\Omega}{2} = \mp \frac{4\sqrt{2} t_c t_{\text{SO}}}{\epsilon + \sqrt{\epsilon^2 + 8t_c^2}},
	\end{align}
where $t_\text{SO} = \sqrt{t_x^2+t_y^2}$. The purely in-plane form of the spin-orbit vector is of particular interest as it is predicted, e.g., for HHs in planar Ge due to the cubic Rashba SOI~\cite{Mutter2021natural} and for holes in Ge/Si nanowires~\cite{Froning2021}. Note that in contrast to the longitudinal case discussed above, the spin-orbit switch can be operated electrically via the parameter $\Omega = \Omega(\epsilon)$, suggesting the possibility of fast and accurate control over the SOI. We display the form of the coupling for exemplary values in Fig.~\figref[(b)]{D0_Dminus}. 

	\begin{figure}
		\includegraphics[scale=0.19]{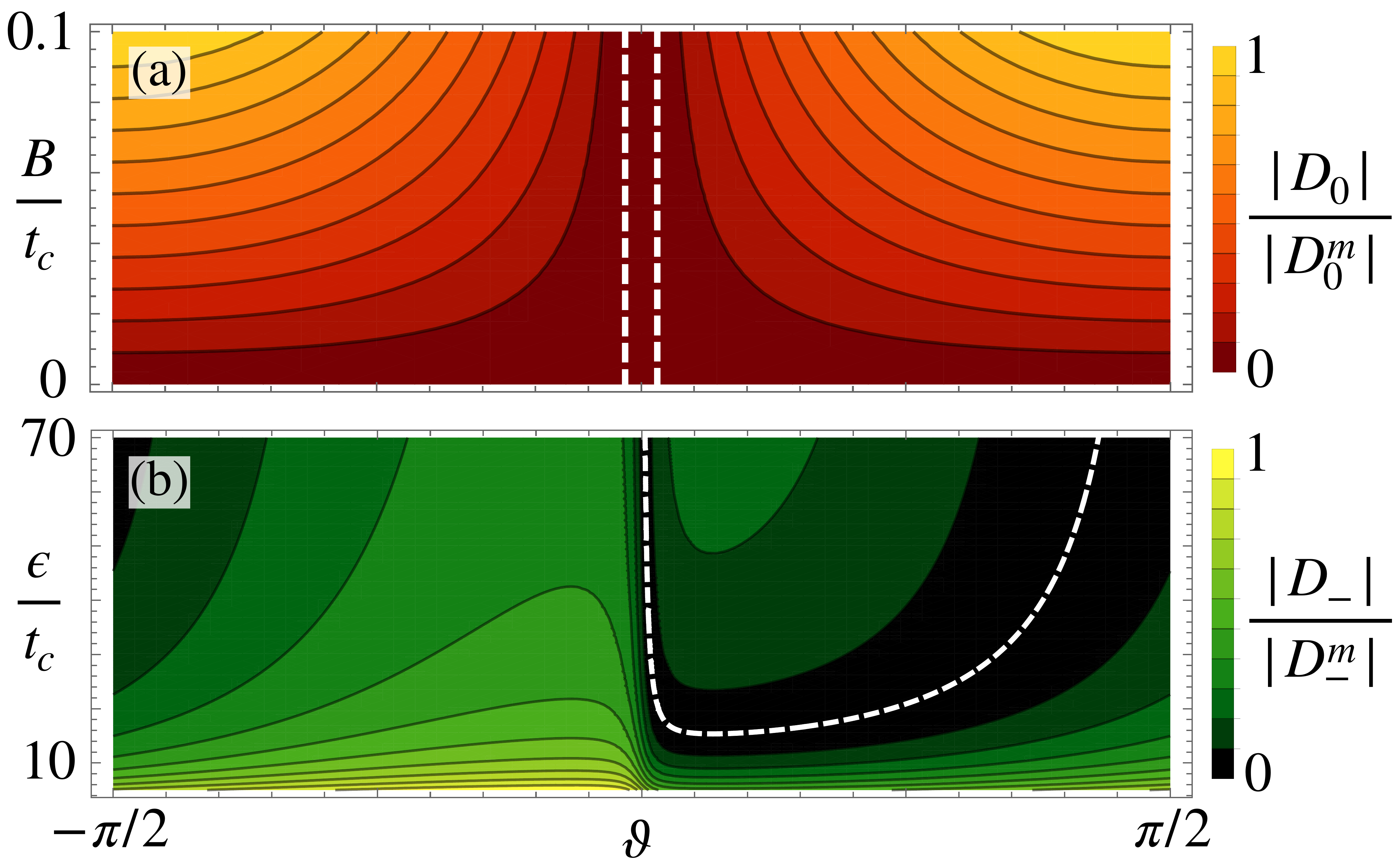}
		\caption{Singlet-triplet coupling. (a) The coupling $D_0$ between the ground-state singlet $\vert S_g \rangle$ and the unpolarized triplet $\vert \mathfrak{T}_0 \rangle$ as a function of the magnitude of the magnetic field $B$ and the tilting angle $\vartheta$ at $\epsilon = 50 t_c$. (b) The coupling $D_-$ between the ground-state singlet $\vert S_g \rangle$ and the polarized triplet $\vert \mathfrak{T}_- \rangle$ as a function of the detuning $\epsilon$ and the tilting angle $\vartheta$ at $B = 0.1 t_c$. One observes zeros for different combinations of the two variables as indicated by the dashed white lines which are drawn according to (a) Eq.~\eqref{eq:zero_line_d0} and (b) Eq.~\eqref{eq:Dpm_ty}. We set $\varphi = -\pi/4$, $\mathbf{t}_{\text{SO}} = (0.05,0.05,0)t_c$ and normalize both couplings by their maximum values $\vert D_0^m \vert$ and $\vert D_-^m \vert$ in the regime displayed.}
		\label{fig:D0_Dminus}
	\end{figure}

On the grounds of symmetry considerations, it is in principle possible that the in-plane degeneracy of the $g$ tensor in HH systems can be lifted if the confinement potential in the dots is elliptical. The broken in-plane symmetry allows for HH-LH induced corrections to the $g$ factors yielding $g^d_x \neq g^d_y$. However, this will leave the above results unchanged if the magnetic field is applied along one of the principal axes of the elliptical confinement potential, $\varphi = 0$ or $\varphi = \pi/2$.

Finally, we point out another application of our results, which is the reverse argumentation of before. By studying the avoided crossing between the ground state singlet and one of the triplet states, one may obtain information about the SOI in the system if the $g$ tensors are known, e.g., from magneto-transport measurements~\cite{Froning2020b, Zhang2021}. The experimental parameters used when the crossing vanishes, e.g. the detuning and magnetic field settings, allow for a determination of the spin-orbit vector. If the $g$ tensors and the spin-orbit vector are known, one may estimate the effect of the nuclear spin bath. Similar schemes have been proposed for systems with equal and isotropic $g$ tensors but non-negligible hyperfine interactions~\cite{Stepanenko2012} and in-plane magnetic fields~\cite{Nowak2011}.

\subsection{Effective qubit Hamiltonians}

Having analyzed the form of the longitudinal and transverse coupling terms, we proceed to study two possible qubit realizations that can benefit from the dependence of the couplings on the magnetic field direction and the detuning described above. First, we consider a $S_g$-$\mathfrak{T}_0$ qubit, i.e., a qubit defined by the ground state singlet $\vert S_g \rangle$ and the unpolarized mixed triplet $\vert \mathfrak{T}_0 \rangle$. By projecting the total Hamiltonian~\eqref{eq:H_hybridized_basis} onto the logical basis, we find the effective qubit Hamiltonian
    \begin{align}
    \label{eq:H_ST0}
        H_{S_g \mathfrak{T}_0} = - J \vert S_g \rangle \langle S_g \vert + \left( D_0 \vert S_g \rangle \langle \mathfrak{T}_0 \vert +  \text{H.c.} \right).
    \end{align}
Since the logical subspace is not ideally decoupled, corrections due to mixing with states outside the qubit space arise. Computing these corrections by means of a Schrieffer-Wolff transformation, however, we find that they are negligible as long as $ \vert D_{\pm} \vert \ll B G_+$ (Appendix~\ref{appx:effective_H_and_leakage}). Moreover, the quality of the approximation increases and leakage errors can be avoided by fixing the magnetic field direction such that $D_- =0$ [cf. Eqs.~\eqref{eq:zero_line_Dpm} and~\eqref{eq:theta_of_phi}]. The coupling $D_0$ can then be switched on and off via electrical $g$ factor engineering~\cite{Nakaoka2007, Prechtel2015, Voisin2016} [cf. Eq.~\eqref{eq:zero_line_d0}] or by controlling the spin-flip tunneling elements [cf. Eq.~\eqref{eq:zero_line_d0_phi}] which are expected to depend on the gate voltages through the applied potential~\cite{Mutter2021natural}.

More direct control via the detuning $\epsilon$ is possible by defining the qubit via the two lowest lying states $\vert S_g \rangle$ and $\vert \mathfrak{T}_- \rangle$. For $\vert D_0 \vert, \vert D_+ \vert \ll B G_+$ the qubit subspace is well isolated (cf. Fig.~\ref{fig:comparison_ana_num}) and by again projecting the total Hamiltonian~\eqref{eq:H_hybridized_basis} onto the logical basis, we obtain the effective qubit Hamiltonian
    \begin{align}
    \label{eq:H_ST-}
        H_{S_g \mathfrak{T}_-} = \left[\frac{BG_+}{2} - J \right] \vert S_g \rangle \langle S_g \vert + \left( D_- \vert S_g \rangle \langle \mathfrak{T}_- \vert + \text{H.c.} \right),
    \end{align}
up to an irrelevant total shift in energy. For a realistic form of the SOI with $\mathbf{t}_{\text{SO}} = (t_x,t_y,0)$ as discussed above, one may set $D_0 = 0$ by fixing the magnetic field such that $ \tan \varphi = - t_x/t_y$ and $\vartheta$ takes the value specified in Eq.~\eqref{eq:zero_line_d0}, resulting in a strong suppression of leakage errors (see Appendix~\ref{appx:effective_H_and_leakage} and Fig.~\figref[(c)]{leakage_via_SWT} therein). By Eq.~\eqref{eq:Dpm_ty}, one then has control over the coupling $D_-$, which is a purely real quantity at the optimal value of $\varphi$, by varying the detuning $\epsilon$, in particular the coupling can be switched on and off electrically. This supplies us with an electrically tunable two-axes control over the qubit. When the detuning is chosen such that $BG_+ = 2 J$ ($D_-=0$), the Hamiltonian generates $x$-rotations ($z$-rotations), and the specific values for $\epsilon_x$ ($\epsilon_z$) can be obtained from Eq.~\eqref{eq:H_hybridized_basis} and Eq.~\eqref{eq:Dpm_ty}, respectively,
    \begin{align}
    \label{eq:detunings}
        &\epsilon_x = \frac{8t_c^2 - (BG_+)^2}{2BG_+}, \\
        &\epsilon_z = \frac{t_c}{t_{\text{SO}}} \frac{  4 t_{\text{SO}}^2 - (BG_a)^2}{\sqrt{2} B G_a}.
    \end{align}
We display the dependence of the elements in the $S_g$-$\mathfrak{T}_-$ qubit Hamiltonian~\eqref{eq:H_ST-} on the detuning in a region including $\epsilon_x$ and $\epsilon_z$ in Fig.~\ref{fig:two_axes_control}. In the same figure we show the Bloch sphere corresponding to the effective two-level system and the orthogonal rotation axes obtained for the two different values of the detuning. We fix the magnetic field at an optimal point by choosing $\varphi$ and $\vartheta$ such that $D_0 = 0$ [cf. Eqs.~\eqref{eq:zero_line_d0} and~\eqref{eq:zero_line_d0_phi}], identify $Z(\epsilon) = J - BG_+/2$ and $X(\epsilon) = D_- $, and set $B= 0.1 t_c$ and $\mathbf{t}_{\text{SO}} = (0.05,0.05,0)t_c$. Note that even at the relatively low magnetic field $B = 0.1 t_c \sim 10$~mT leakage to the only relevant coupled state $\mathfrak{T}_+$ is negligible (Appendix~\ref{appx:effective_H_and_leakage}). At this point, one has $Z(\epsilon_z) \approx 0.06 t_c $ and $X(\epsilon_x) = 0.003 t_c$, allowing for Rabi $z$-rotations in the GHz range and $x$-rotations at tens of MHz. Hence, it is feasible to operate the system below the critical field of conventional superconductors such as aluminum, suggesting that all-electrically controllable, fast $S_g$-$\mathfrak{T}_-$ qubits can be coupled using superconducting transmission lines~\cite{Burkard2006,Jin2012}.

\section{Strong spin-orbit interaction}
\label{sec:SO_switch_strong_SOI}

For systems where the elements of the spin-orbit vector are comparable in magnitude to the tunnel coupling $t_c$, the states defined in Eq.~\eqref{eq:hybridized_basis} will no longer provide an appropriate basis. Moreover, since low magnetic fields are attractive considering integration of the qubit with superconducting resonators, it is natural to study the regime $ B G_+ \ll \vert \mathbf{t}_{\text{SO}} \vert \equiv t_{\text{SO}}$. One may then exactly diagonalize the dominant part of the Hamiltonian $H_0 + H_{\text{SOI}}$ and treat the Zeeman term $H_Z$ as a small perturbation. We find three degenerate states $\vert \tilde{T}_0 \rangle$, $\vert \tilde{T}_+ \rangle$, $\vert \tilde{T}_- \rangle$ mixing all (1,1) states with zero energy and two states $\vert \pm \rangle$ that also contain an admixture of the (2,0) singlet with energies
	\begin{align}
	\label{eq:energies_SOI}
		E_{\pm} = \frac{\epsilon}{2} \pm \sqrt{\frac{\epsilon^2}{4} + 2 t_c^2 + 2 t^2_{\text{SO}} }.
	\end{align}
The exact form of the eigenstates can be found in Appendix~\ref{appx:eigenbasis_SOI}. We may identify the states $\vert + \rangle  \leftrightarrow \vert S_e \rangle$ and  $\vert - \rangle  \leftrightarrow \vert S_g \rangle$ in the sense that they will transition into each other for $t_{\text{SO}} \rightarrow 0$. At zero magnetic field the states $\vert \tilde{T}_{\nu} \rangle$ with $\nu \in \lbrace 0, \pm \rbrace$ form an eigenspace with zero energy. Looking at their behaviour at finite field as $t_{\text{SO}}$ tends to zero we may identify $\vert \tilde{T}_{\nu} \rangle \leftrightarrow \vert T_{\nu} \rangle$. Hence, if we are interested in singlet-triplet like mixing, the relevant couplings are induced by the Zeeman term,
	\begin{align}
	\label{eq:couplings_strong_SOI}
		C_{\nu} \equiv   \langle - \vert H_Z \vert \tilde{T}_{\nu} \rangle = B \cdot f_{\nu} \left(g_x^{\pm}, g_z^{\pm}, t_x,t_y,t_z \right),
	\end{align}
where $\nu \in \lbrace 0, \pm \rbrace$ and the functions $f_{\nu}$ contain all information about anisotropic site-dependent $g$ tensors and the SOI. As a consequence, the non-trivial zeros of $C_{\nu}$ are determined by the orientation of the magnetic field alone. We note that as in Sec.~\ref{sec:spin-orbit_switch} the expressions are unchanged when we decouple the energetically high lying state $\vert + \rangle$ via a Schrieffer-Wolff transformation to leading order in $B G_+/\Delta E$ where the denominator denotes the energy separation $\Delta E \sim \sqrt{\epsilon^2 + 8 t_c^2 + 8 t^2_{\text{SO}}}$ between the states in the subspaces and $B G_+$ is the relevant scale of the mixing terms. 

Clearly, in an eigenbasis of $H_0 + H_{\text{SOI}}$, the SOI does not amount to a coupling of the eigenstates. One may then control the couplings simply by tuning the magnitude of the magnetic field. However, for gate operation it is desirable to have the coupling between the logical qubit states (e.g., $C_0$ turned on), while all other couplings should be absent to avoid leakage errors. This can be achieved by controlling the orientation of the magnetic field relative to the spin-orbit vector. The explicit form of the couplings $C_{\nu}$ is easily computed from Eq.~\eqref{eq:couplings_strong_SOI} but lengthy, and we restrict our attention to determining the the points where $C_{\nu} = 0$.

	\begin{figure}
		\includegraphics[scale=0.305]{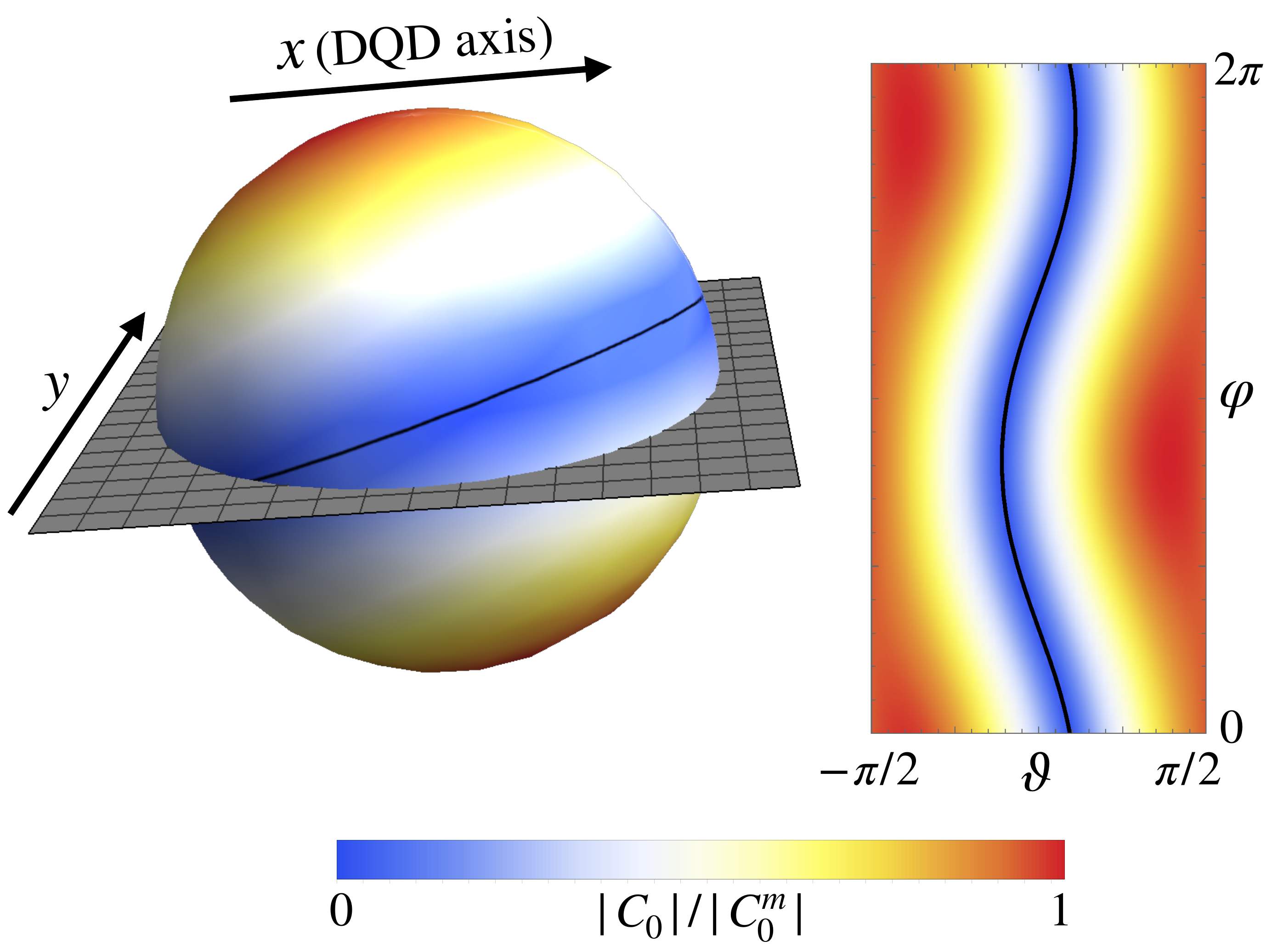}
		\caption{The coupling $C_0$ as a function of the magnetic field direction. We display a spherical plot with inset $x$-$y$-plane (left) and the corresponding planar plot as a function of the polar angle $\vartheta$ and the azimuthal angle $\varphi $ (right). The line of zero coupling (solid black line) is drawn according to Eq.~\eqref{eq:C0_zero}. We set $\epsilon = 100t_c$, $\mathbf{t}_{\text{SO}} = (t_x, t_y, t_z) = (2,2,1)t_c$ and normalize the coupling by its maximum value $\vert C_0^m \vert$ in the regime under consideration.}
		\label{fig:C0_angles}
	\end{figure}

One finds $C_0 = 0$ for any azimuthal angle $\varphi$ as long as the polar angle is chosen such that
	\begin{align}
	\begin{split}
	\label{eq:C0_zero}
		& \tan \vartheta =   \frac{g_x^+ t_y t_c - g_x^- t_x t_z}{g_z^-(t_c^2 + t_z^2)} \cos \varphi - \frac{g_x^+ t_x t_c + g_x^- t_y t_z}{g_z^-(t_c^2 + t_z^2)} \sin \varphi. 
	\end{split}
	\end{align}
We display the absolute value of $C_0$ and its dependence on the magnetic field direction for exemplary values of the spin-orbit vector in Fig.~\ref{fig:C0_angles}. Moreover, we find $C_+ = 0$ if the following conditions are satisfied,
	\begin{align}
	\label{eq:C+_zero}
	\begin{split}
		& g_z^+t_x (t_c^2 + t_z^2)\sin \vartheta  = \xi_1 (\varphi) \cos \vartheta, \\
		&  \xi_1 (\varphi) = \left(g_x^- t_c t_x t_y + g_x^+t_z (t_c^2+t_y^2+t_z^2) \right) \cos \varphi  \\
		& \qquad \quad + \left(g_x^- t_c (t_c^2+t_y^2+t_z^2) - g_x^+ t_x t_y t_z \right) \sin \varphi, \\
		& g_z^+t_y (t_c^2 + t_z^2)\sin \vartheta = \xi_2(\varphi) \cos \vartheta, \\
		& \xi_2(\varphi) =  -\left(g_x^+ t_x t_y t_z + g_x^- t_c ( t_c^2 + t_x^2 + t_z^2) \right) \cos \varphi \\
		&\qquad \quad  +  \left(g_x^+ t_z ( t_c^2+t_x^2+t_z^2) - g_x^- t_x t_y t_c \right) \sin \varphi,
	\end{split}
	\end{align}
and $C_- = 0$ for
	\begin{align}
	\label{eq:C-_zero}
	\begin{split}
		& g_z^+ t_x \sin \vartheta = \left( g_x^+ t_z \cos \varphi +g_x^- t_c \sin \varphi \right) \cos \vartheta, \\
		& g_z^+ t_y \sin \vartheta = \left( g_x^+ t_z \sin \varphi - g_x^- t_c \cos\varphi \right) \cos \vartheta.
	\end{split}
	\end{align}
For a given spin-orbit vector, the conditions~\eqref{eq:C0_zero}-\eqref{eq:C-_zero} can easily be rearranged to obtain the angles $\vartheta $ and $\varphi$ solely in terms of the tunnel coupling, the spin-orbit vector and the $g$ factors.

Finally, we consider the special case of an in-plane spin-orbit vector $\mathbf{t}_{\text{SO}} = (t_x,t_y,0)$ as in Sec.~\ref{sec:spin-orbit_switch}. We encode the qubit in the states $ \vert - \rangle$ and $ \vert \tilde{T}_0 \rangle$, which are split by the modified exchange energy,
    \begin{align}
    \label{eq:extended_exchange_energy}
        \tilde{J} = \sqrt{\frac{\epsilon^2}{4} + 2 t_c^2 + 2 t^2_{\text{SO}}} - \frac{\epsilon }{2} > 0,
    \end{align}
at zero magnetic field. As can be seen from Eqs.~\eqref{eq:C+_zero} and~\eqref{eq:C-_zero}, we may set $C_+ = C_- = 0$ by choosing
	\begin{align}
	\label{eq:zero_angles_HHs}
		\varphi = - \arctan \left( \frac{t_x}{t_y} \right), \quad  \vartheta = - \arctan \left( \frac{g_x^- t_c }{g_z^+ t_{\text{SO}}} \right).
	\end{align}
For these values, one has $C_0 \neq 0 $ and so the qubit ground state $\vert - \rangle$ is completely decoupled from the states outside the logical subspace but coupled to the excited qubit state; the strength of the coupling can be controlled by the magnitude of the magnetic field $B$. Of course, the coupling cannot be made arbitrarily large in this way since the condition $BG_+ \ll t_{\text{SO}}$ must be met to ensure that the approach presented in this section can accurately describe the physical system.

\section{Conclusion}
\label{sec:conclusion}

Starting from a general phenomenological model for spins confined in a DQD system subject to a magnetic field, we show the existence of magnetic field strengths and directions where the coupling between the ground state singlet and the triplet states vanishes. The effect is due to cancelling contributions from the SOI and anisotropic site-dependent $g$ tensors, both of which are characteristic features of semiconductor hole systems. With the magnetic field fixed at one such optimal point one may avoid leakage to states outside the logical subspace and at the same time operate the qubit all-electrically. In particular, we demonstrate that for realistic system parameters an electric spin-orbit switch can be implemented and controlled by the detuning, allowing for highly tunable two-axes control in hole singlet-triplet qubits. This property can be used in quantum information processing units built from singlet-triplet qubits to tune into regimes best suited for the tasks of readout and initialization (low coupling) as well as qubit manipulation (high coupling). Since fast operations at optimal points are feasible at low magnetic fields ($\sim 10$~mT), different singlet-triplet qubits can be interconnected using superconducting resonators.

Moreover, we study systems with a strong SOI operated at low magnetic fields, and demonstrate that for a realistic form of the spin-orbit vector there exists a particular direction of the magnetic field where the coupling between the qubit ground state and states outside the logical subspace can be set to zero. At this optimal point the qubit ground state is protected from leakage errors, and the strength of the coupling to the excited qubit state can be tuned by the magnitude of the magnetic field.

Future research may focus on the precise form of the SOI, e.g., in two-dimensional Ge HH systems to include effects such as a magnetic field dependence of the spin-orbit elements as well as spin-orbit terms induced by magnetic fields which are not time reversal invariant and thus not described by the model used in this work. 

\section{Acknowledgements}
We would like to thank Daniel Jirovec and Georgios Katsaros for insightful discussions on hole systems in planar germanium. This research is supported by the German Research Foundation (Deutsche Forschungsgemeinschaft, DFG) under project number 450396347.

\appendix

\section{Fixing the quantization axis}
\label{appx:quantization_axis}

In this appendix we relate the results of the main text to the case where one may choose the orientation of the quantization axis freely. Defining the spin operators and the effective magnetic field in dot $d$ as $S_i^d = \sigma_i^d/2$ and $\mathfrak{B}_i^d = \sum_j g^d_{ij}B_j$, respectively, one finds for the Zeeman Hamiltonian,
	\begin{align}
	\begin{split}
		& H_Z  =  \frac{1}{2}  \boldsymbol{\mathcal{B}}^+\cdot   \left( \mathbf{S}^L + \mathbf{S}^R \right) + \frac{1}{2}  \boldsymbol{\mathcal{B}}^- \cdot  \left( \mathbf{S}^L - \mathbf{S}^R \right),
	\end{split}
	\end{align}
where we define $ \boldsymbol{\mathcal{B}}^{\pm} =  \boldsymbol{\mathcal{B}}^L \pm \boldsymbol{\mathcal{B}}^R$. Choosing the quantization axis along $\boldsymbol{\mathcal{B}}^+$ and denoting this direction by $x_3$, we find by decomposing $\boldsymbol{\mathcal{B}}^-$ into components parallel and perpendicular to $\boldsymbol{\mathcal{B}}^+$,
	\begin{align}
	\label{eq:H_Z_quantization_axis}
	\begin{split}
		&H_Z = \frac{BG_+}{2} \left( S_3^L + S_3^R \right) + \frac{BG_-}{2} \left( S_3^L - S_3^R \right)  \\
		& \qquad + \frac{B G_a}{\sqrt{2}} \left[ \cos \xi \left( S_1^L - S_1^R \right) + \sin \xi \left( S_2^L - S_2^R \right) \right],
	\end{split}
	\end{align}
where the effective $g$ factors are as given in Eq.~\eqref{eq:effective_g_factors} and $S_i$ is the spin operator along $x_i$, $i =1,2,3$. The angle $\xi$ only amounts to an irrelevant phase and may thus be set to zero. One can see from the Hamiltonian~\eqref{eq:H_Z_quantization_axis} that the hybridized triplets~\eqref{eq:hybridized_basis} are the same as those obtained by choosing the quantization axis along the sum of effective magnetic fields.

While the form of the part of the Hamiltonian describing the detuning and tunneling between the singlets is invariant under a change of the quantization axis, the part of the Hamiltonian describing the SOI may be transformed into the basis defined by the above choice of the quantization axis, and we recover the matrix elements in Eq.~\eqref{eq:matrix_elements_SOI} after performing an additional transformation into the basis of hybridized singlets [cf. Eq.~\eqref{eq:hybridized_basis}].

\section{Effective qubit Hamiltonians and leakage}
\label{appx:effective_H_and_leakage}

In this appendix we quantify our claim of Sec.~\ref{sec:spin-orbit_switch} and show that the effective qubit Hamiltonians shown in Eqs.~\eqref{eq:H_ST0} and~\eqref{eq:H_ST-} are accurate since leakage to states outside the logical subspace is suppressed. For this purpose we decouple the qubit space from the remaining states in the (1,1) charge configuration by means of a Schrieffer-Wolff transformation.

For the Hamiltonian in Eq.~\eqref{eq:H_ST0}, the logical subspace is spanned by the states $\vert S_g \rangle$ and $\vert \mathfrak{T}_0 \rangle$. Upon application of a Schrieffer-Wolff transformation, these states will be changed to the perturbed states $\vert S_g^{\text{eff}} \rangle$ and $\vert \mathfrak{T}_0^{\text{eff}} \rangle$, and we find to leading order,
    \begin{align}
    \label{eq:effective_HST0_plus_corrections}
    \begin{split}
        &H^{\text{eff}}_{S_g \mathfrak{T}_0} = \left(- J + BG_+[ \Delta_+ - \Delta_-] \right) \vert S_g^{\text{eff}} \rangle \langle S_g^{\text{eff}} \vert  \\
        & \qquad \quad + \left( D_0 [1+ \Delta_+ + \Delta_-] \vert S_g^{\text{eff}} \rangle \langle \mathfrak{T}_0^{\text{eff}} \vert + \text{H.c.} \right),  
    \end{split}
    \end{align}
where
    \begin{align}
        & \Delta_{\pm} = \frac{2 \vert D_{\pm} \vert^2 }{4 \vert D_0 \vert^2 \mp BG_+ (2 J \pm BG_+)}.
    \end{align}
At $\Delta_+ = \Delta_- = 0$ we recover Eq.~\eqref{eq:H_ST0} of the main text. Working at an optimal point, we can achieve $\Delta_- =0$ by fixing the direction of the magnetic field or the detuning according to Eqs.~\eqref{eq:zero_line_Dpm} and~\eqref{eq:theta_of_phi}. The term $\Delta_+$ then determines the remaining leakage out of the original logical subspace $\lbrace \vert S_g \rangle$, $\vert \mathfrak{T}_0 \rangle \rbrace$ to the state $\vert \mathfrak{T}_+ \rangle$.

An analogous analysis applies for the case of the $S_g$-$\mathfrak{T}_-$ qubit Hamiltonian~\eqref{eq:H_ST-}, yielding
    \begin{align}
    \label{eq:effective_HST-_plus_corrections}
    \begin{split}
        &H^{\text{eff}}_{S_g \mathfrak{T}_-} = \left( \frac{BG_+}{2}[1 + 2 \delta_0 + 4 \delta_+] - J \right) \vert S_g^{\text{eff}}\rangle \langle S_g^{\text{eff}} \vert  \\
        & \qquad \qquad + \left( D_-[1+ \delta_0 + \delta_+] \vert S_g^{\text{eff}} \rangle \langle \mathfrak{T}_-^{\text{eff}} \vert + \text{H.c.} \right),  
    \end{split}
    \end{align}
where
    \begin{align}
        & \delta_0 = \frac{\vert D_0 \vert^2 }{2 \vert D_- \vert^2 - B G_+ J}, \\
        &  \delta_+ = \frac{\vert D_+ \vert^2 }{2 \vert D_- \vert^2 - B G_+(BG_+ +2 J)}.
    \end{align}
One may achieve $\delta_0 =0$ by choosing the magnetic field orientation according to Eqs.~\eqref{eq:zero_line_d0} and~\eqref{eq:zero_line_d0_phi}. At this point the term $\delta_+$ determines the residual leakage out of the original logical subspace $\lbrace \vert S_g \rangle$, $\vert \mathfrak{T}_- \rangle \rbrace$ to the state $\vert \mathfrak{T}_+ \rangle$.

\begin{figure*}
    \includegraphics[scale=0.28]{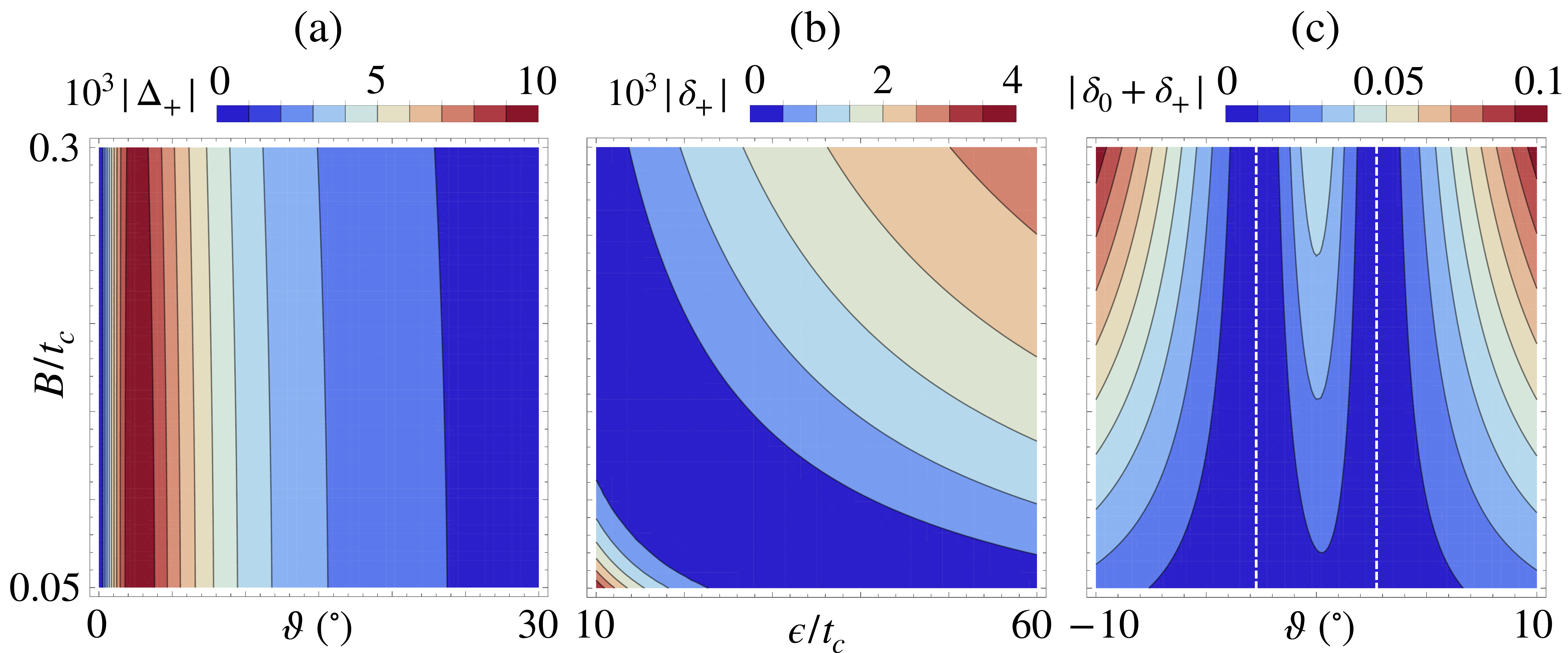}
    \caption{Leakage in hole singlet-triplet qubits. (a) $S_g$-$\mathfrak{T}_0$ qubit. Leakage $\vert \Delta_+ \vert$ to the state $\vert \mathfrak{T}_+ \rangle$ as a function of the magnetic field strength and its polar angle at the optimal point $\Delta_-=0$. One finds $\vert \Delta_+ \vert < 0.001 $ for all $\vartheta$ outside the region shown. The magnetic field direction can be fixed to a region of low leakage (blue) by fixing the detuning [Eq.~\eqref{eq:detunings}]. (b),(c) $S_g$-$\mathfrak{T}_-$ qubit. (b) Leakage $\vert \delta_+ \vert$ to the state $\vert \mathfrak{T}_+ \rangle$ as a function of the magnetic field strength and the detuning at an optimal point where $D_0 = 0$.  One finds $\vert \delta_+ \vert <0.002$  even for magnetic fields $B$ as low as $0.1 t_c$. (c) The total leakage $ \vert \delta_0 + \delta_+ \vert$ to the states $\vert \mathfrak{T}_0 \rangle$ and $\vert \mathfrak{T}_+ \rangle$ at $\epsilon = 50 t_c$ in a region around the optimal points at $\vartheta = \pm \arctan \sqrt{-g_x^-g_x^+/g_z^-g_z^+}$ (dashed white lines). Sizeable leakage errors arise when the magnetic field orientation is not chosen such that $D_0 = 0$. The leakage is even larger for angles outside the region shown. We set $\mathbf{t}_{\text{SO}} = (0.05,0.05,0)t_c$ and $\varphi = - \pi/4$.}
    \label{fig:leakage_via_SWT}
\end{figure*}

As can be seen from Fig.~\ref{fig:leakage_via_SWT}, the leakage for both qubits discussed above is well below one percent for realistic parameter settings. For the $S_g$-$\mathfrak{T}_0$ qubit, the detuning can be chosen such that $\vert \Delta_+ \vert < 0.001$. On the other hand, for the $S_g$-$\mathfrak{T}_-$ qubit, one has $\vert \delta_- \vert < 0.002$ at $B=0.1 t_c$ for the detuning values required for gate operation (cf. Fig.~\ref{fig:two_axes_control}). Fig.~\figref[(c)]{leakage_via_SWT} highlights the strong suppression of leakage in the $S_g$-$\mathfrak{T}_-$ qubit at an optimal point with $D_0 = 0$, emphasizing the importance of operating the singlet-triplet qubit at the optimal points reported in this paper. 

\section{Eigenbasis of the orbital and spin-orbit Hamiltonian}
\label{appx:eigenbasis_SOI}
The Hamiltonian $H_0 + H_{\text{SO}}$ given in Eq.~\eqref{eq:model_Hamiltonian} of the main text may be diagonalized exactly, yielding the orthonormal eigenstates,
    \begin{widetext}
	\begin{align}
	\label{eq:eigenstates:SOI}
		&\vert \tilde{T}_0 \rangle = \frac{1}{\sqrt{t_c^2+t_z^2}} \left( i t_z \vert S \rangle + t_c \vert T_0 \rangle \right), \\
		& \vert \tilde{T}_+ \rangle = \frac{1}{\sqrt{2t_c^2(t_c^2+t_z^2)(2t_c^2 + 2t_z^2 +t_x^2 + t_y^2)}} \left( \sqrt{2}(t_y + it_x)t_c^2 \vert S \rangle - i \sqrt{2}(t_y + it_x)t_c t_z \vert T_0 \rangle  + 2t_c(t_c^2+t_z^2) \vert T_+ \rangle  \right), \\
		& \vert \tilde{T}_- \rangle =  \frac{\sqrt{8}(t_y - it_x)t_c(t_c^2+t_z^2) [ t_c \vert S \rangle -it_z \vert T_0 \rangle ]   -2 t_c(t_c^2+ t_z^2)(t_y - i t_x)^2 \vert T_+ \rangle + 2t_c (t_c^2 + t_z^2) (2t_c^2 + 2t_z^2 + t_x^2 + t_y^2) \vert T_- \rangle }{\sqrt{8t_c^2(t_c^2+t_z^2)^2(t_c^2+ t_{\text{SO}}^2)(2t_c^2 + 2t_z^2 +t_x^2 + t_y^2)} }, \\
		& \vert \pm \rangle = \frac{1}{\sqrt{4(t_c^2+t_{\text{SO}}^2) + \epsilon E_{\pm} }} \left( i E_{\pm} \vert S_{20} \rangle + \sqrt{2} i t_c \vert S \rangle + \sqrt{2} t_z \vert T_0 \rangle - (t_x + it_y) \vert T_+ \rangle + (t_x - i t_y) \vert T_- \rangle \right),
	\end{align}
	\end{widetext}
where $t_{\text{SO}} = \vert \mathbf{t}_{\text{SO}} \vert$.
The states $\vert \tilde{T}_0 \rangle$, $\vert \tilde{T}_+ \rangle$, $\vert \tilde{T}_- \rangle$ are degenerate with zero energy and have been orthogonalized by hand, while the states~$\vert \pm \rangle$ have energies
	\begin{align}
	\label{eq:SOI_energies}
		E_{\pm} = \frac{\epsilon}{2} \pm \sqrt{\frac{\epsilon^2}{4} + 2 t_c^2 + 2 t^2_{\text{SO}} }.
	\end{align}

\bibliography{Ge_holes_literature}

\begin{thebibliography}{53}%
\makeatletter
\providecommand \@ifxundefined [1]{%
 \@ifx{#1\undefined}
}%
\providecommand \@ifnum [1]{%
 \ifnum #1\expandafter \@firstoftwo
 \else \expandafter \@secondoftwo
 \fi
}%
\providecommand \@ifx [1]{%
 \ifx #1\expandafter \@firstoftwo
 \else \expandafter \@secondoftwo
 \fi
}%
\providecommand \natexlab [1]{#1}%
\providecommand \enquote  [1]{``#1''}%
\providecommand \bibnamefont  [1]{#1}%
\providecommand \bibfnamefont [1]{#1}%
\providecommand \citenamefont [1]{#1}%
\providecommand \href@noop [0]{\@secondoftwo}%
\providecommand \href [0]{\begingroup \@sanitize@url \@href}%
\providecommand \@href[1]{\@@startlink{#1}\@@href}%
\providecommand \@@href[1]{\endgroup#1\@@endlink}%
\providecommand \@sanitize@url [0]{\catcode `\\12\catcode `\$12\catcode
  `\&12\catcode `\#12\catcode `\^12\catcode `\_12\catcode `\%12\relax}%
\providecommand \@@startlink[1]{}%
\providecommand \@@endlink[0]{}%
\providecommand \url  [0]{\begingroup\@sanitize@url \@url }%
\providecommand \@url [1]{\endgroup\@href {#1}{\urlprefix }}%
\providecommand \urlprefix  [0]{URL }%
\providecommand \Eprint [0]{\href }%
\providecommand \doibase [0]{https://doi.org/}%
\providecommand \selectlanguage [0]{\@gobble}%
\providecommand \bibinfo  [0]{\@secondoftwo}%
\providecommand \bibfield  [0]{\@secondoftwo}%
\providecommand \translation [1]{[#1]}%
\providecommand \BibitemOpen [0]{}%
\providecommand \bibitemStop [0]{}%
\providecommand \bibitemNoStop [0]{.\EOS\space}%
\providecommand \EOS [0]{\spacefactor3000\relax}%
\providecommand \BibitemShut  [1]{\csname bibitem#1\endcsname}%
\let\auto@bib@innerbib\@empty
\bibitem [{\citenamefont {Hendrickx}\ \emph
  {et~al.}(2020{\natexlab{a}})\citenamefont {Hendrickx}, \citenamefont
  {Franke}, \citenamefont {Sammak}, \citenamefont {Scappucci},\ and\
  \citenamefont {Veldhorst}}]{Hendrickx2020}%
  \BibitemOpen
  \bibfield  {author} {\bibinfo {author} {\bibfnamefont {N.}~\bibnamefont
  {Hendrickx}}, \bibinfo {author} {\bibfnamefont {D.}~\bibnamefont {Franke}},
  \bibinfo {author} {\bibfnamefont {A.}~\bibnamefont {Sammak}}, \bibinfo
  {author} {\bibfnamefont {G.}~\bibnamefont {Scappucci}},\ and\ \bibinfo
  {author} {\bibfnamefont {M.}~\bibnamefont {Veldhorst}},\ }\bibfield  {title}
  {\bibinfo {title} {Fast two-qubit logic with holes in germanium},\ }\href
  {https://doi.org/10.1038/s41586-019-1919-3} {\bibfield  {journal} {\bibinfo
  {journal} {Nature}\ }\textbf {\bibinfo {volume} {577}},\ \bibinfo {pages}
  {487} (\bibinfo {year} {2020}{\natexlab{a}})}\BibitemShut {NoStop}%
\bibitem [{\citenamefont {Hendrickx}\ \emph
  {et~al.}(2020{\natexlab{b}})\citenamefont {Hendrickx}, \citenamefont
  {Lawrie}, \citenamefont {Petit}, \citenamefont {Sammak}, \citenamefont
  {Scappucci},\ and\ \citenamefont {Veldhorst}}]{Hendrickx2020b}%
  \BibitemOpen
  \bibfield  {author} {\bibinfo {author} {\bibfnamefont {N.}~\bibnamefont
  {Hendrickx}}, \bibinfo {author} {\bibfnamefont {W.}~\bibnamefont {Lawrie}},
  \bibinfo {author} {\bibfnamefont {L.}~\bibnamefont {Petit}}, \bibinfo
  {author} {\bibfnamefont {A.}~\bibnamefont {Sammak}}, \bibinfo {author}
  {\bibfnamefont {G.}~\bibnamefont {Scappucci}},\ and\ \bibinfo {author}
  {\bibfnamefont {M.}~\bibnamefont {Veldhorst}},\ }\bibfield  {title} {\bibinfo
  {title} {A single-hole spin qubit},\ }\href
  {https://doi.org/10.1038/s41467-020-17211-7} {\bibfield  {journal} {\bibinfo
  {journal} {Nat. Commun.}\ }\textbf {\bibinfo {volume} {11}},\ \bibinfo
  {pages} {3478} (\bibinfo {year} {2020}{\natexlab{b}})}\BibitemShut {NoStop}%
\bibitem [{\citenamefont {Wang}\ \emph {et~al.}(2020)\citenamefont {Wang},
  \citenamefont {Xu}, \citenamefont {Gao}, \citenamefont {Liu}, \citenamefont
  {Ma}, \citenamefont {Zhang}, \citenamefont {Zhang}, \citenamefont {Cao},
  \citenamefont {Wang}, \citenamefont {Zhang}, \citenamefont {Hu},
  \citenamefont {Jiang}, \citenamefont {Li}, \citenamefont {Guo},\ and\
  \citenamefont {Guo}}]{Wang2020arXiv}%
  \BibitemOpen
  \bibfield  {author} {\bibinfo {author} {\bibfnamefont {K.}~\bibnamefont
  {Wang}}, \bibinfo {author} {\bibfnamefont {G.}~\bibnamefont {Xu}}, \bibinfo
  {author} {\bibfnamefont {F.}~\bibnamefont {Gao}}, \bibinfo {author}
  {\bibfnamefont {H.}~\bibnamefont {Liu}}, \bibinfo {author} {\bibfnamefont
  {R.-L.}\ \bibnamefont {Ma}}, \bibinfo {author} {\bibfnamefont
  {X.}~\bibnamefont {Zhang}}, \bibinfo {author} {\bibfnamefont
  {T.}~\bibnamefont {Zhang}}, \bibinfo {author} {\bibfnamefont
  {G.}~\bibnamefont {Cao}}, \bibinfo {author} {\bibfnamefont {T.}~\bibnamefont
  {Wang}}, \bibinfo {author} {\bibfnamefont {J.-J.}\ \bibnamefont {Zhang}},
  \bibinfo {author} {\bibfnamefont {X.}~\bibnamefont {Hu}}, \bibinfo {author}
  {\bibfnamefont {H.-W.}\ \bibnamefont {Jiang}}, \bibinfo {author}
  {\bibfnamefont {H.-O.}\ \bibnamefont {Li}}, \bibinfo {author} {\bibfnamefont
  {G.-C.}\ \bibnamefont {Guo}},\ and\ \bibinfo {author} {\bibfnamefont {G.-P.}\
  \bibnamefont {Guo}},\ }\href@noop {} {\bibinfo {title} {Ultrafast operations
  of a hole spin qubit in {Ge} quantum dot}} (\bibinfo {year} {2020}),\ \Eprint
  {https://arxiv.org/abs/2006.12340} {arXiv:2006.12340 [cond-mat.mes-hall]}
  \BibitemShut {NoStop}%
\bibitem [{\citenamefont {Lawrie}\ \emph {et~al.}(2020)\citenamefont {Lawrie},
  \citenamefont {Hendrickx}, \citenamefont {van Riggelen}, \citenamefont
  {Russ}, \citenamefont {Petit}, \citenamefont {Sammak}, \citenamefont
  {Scappucci},\ and\ \citenamefont {Veldhorst}}]{Lawrie2020}%
  \BibitemOpen
  \bibfield  {author} {\bibinfo {author} {\bibfnamefont {W.~I.~L.}\
  \bibnamefont {Lawrie}}, \bibinfo {author} {\bibfnamefont {N.~W.}\
  \bibnamefont {Hendrickx}}, \bibinfo {author} {\bibfnamefont {F.}~\bibnamefont
  {van Riggelen}}, \bibinfo {author} {\bibfnamefont {M.}~\bibnamefont {Russ}},
  \bibinfo {author} {\bibfnamefont {L.}~\bibnamefont {Petit}}, \bibinfo
  {author} {\bibfnamefont {A.}~\bibnamefont {Sammak}}, \bibinfo {author}
  {\bibfnamefont {G.}~\bibnamefont {Scappucci}},\ and\ \bibinfo {author}
  {\bibfnamefont {M.}~\bibnamefont {Veldhorst}},\ }\bibfield  {title} {\bibinfo
  {title} {Spin relaxation benchmarks and individual qubit addressability for
  holes in quantum dots},\ }\href
  {https://doi.org/10.1021/acs.nanolett.0c02589} {\bibfield  {journal}
  {\bibinfo  {journal} {Nano Letters}\ }\textbf {\bibinfo {volume} {20}},\
  \bibinfo {pages} {7237} (\bibinfo {year} {2020})}\BibitemShut {NoStop}%
\bibitem [{\citenamefont {van Riggelen}\ \emph {et~al.}(2021)\citenamefont {van
  Riggelen}, \citenamefont {Hendrickx}, \citenamefont {Lawrie}, \citenamefont
  {Russ}, \citenamefont {Sammak}, \citenamefont {Scappucci},\ and\
  \citenamefont {Veldhorst}}]{van_Riggelen2021}%
  \BibitemOpen
  \bibfield  {author} {\bibinfo {author} {\bibfnamefont {F.}~\bibnamefont {van
  Riggelen}}, \bibinfo {author} {\bibfnamefont {N.~W.}\ \bibnamefont
  {Hendrickx}}, \bibinfo {author} {\bibfnamefont {W.~I.~L.}\ \bibnamefont
  {Lawrie}}, \bibinfo {author} {\bibfnamefont {M.}~\bibnamefont {Russ}},
  \bibinfo {author} {\bibfnamefont {A.}~\bibnamefont {Sammak}}, \bibinfo
  {author} {\bibfnamefont {G.}~\bibnamefont {Scappucci}},\ and\ \bibinfo
  {author} {\bibfnamefont {M.}~\bibnamefont {Veldhorst}},\ }\bibfield  {title}
  {\bibinfo {title} {A two-dimensional array of single-hole quantum dots},\
  }\href {https://doi.org/10.1063/5.0037330} {\bibfield  {journal} {\bibinfo
  {journal} {Applied Physics Letters}\ }\textbf {\bibinfo {volume} {118}},\
  \bibinfo {pages} {044002} (\bibinfo {year} {2021})}\BibitemShut {NoStop}%
\bibitem [{\citenamefont {Hendrickx}\ \emph {et~al.}(2021)\citenamefont
  {Hendrickx}, \citenamefont {Lawrie}, \citenamefont {Russ}, \citenamefont {van
  Riggelen}, \citenamefont {de~Snoo}, \citenamefont {Schouten}, \citenamefont
  {Sammak}, \citenamefont {Scappucci},\ and\ \citenamefont
  {Veldhorst}}]{Hendrickx2021four}%
  \BibitemOpen
  \bibfield  {author} {\bibinfo {author} {\bibfnamefont {N.~W.}\ \bibnamefont
  {Hendrickx}}, \bibinfo {author} {\bibfnamefont {W.~I.}\ \bibnamefont
  {Lawrie}}, \bibinfo {author} {\bibfnamefont {M.}~\bibnamefont {Russ}},
  \bibinfo {author} {\bibfnamefont {F.}~\bibnamefont {van Riggelen}}, \bibinfo
  {author} {\bibfnamefont {S.~L.}\ \bibnamefont {de~Snoo}}, \bibinfo {author}
  {\bibfnamefont {R.~N.}\ \bibnamefont {Schouten}}, \bibinfo {author}
  {\bibfnamefont {A.}~\bibnamefont {Sammak}}, \bibinfo {author} {\bibfnamefont
  {G.}~\bibnamefont {Scappucci}},\ and\ \bibinfo {author} {\bibfnamefont
  {M.}~\bibnamefont {Veldhorst}},\ }\bibfield  {title} {\bibinfo {title} {A
  four-qubit germanium quantum processor},\ }\href
  {https://doi.org/10.1038/s41586-021-03332-6} {\bibfield  {journal} {\bibinfo
  {journal} {Nature}\ }\textbf {\bibinfo {volume} {591}},\ \bibinfo {pages}
  {580} (\bibinfo {year} {2021})}\BibitemShut {NoStop}%
\bibitem [{\citenamefont {Wang}\ \emph {et~al.}(2021)\citenamefont {Wang},
  \citenamefont {Marcellina}, \citenamefont {Hamilton}, \citenamefont {Cullen},
  \citenamefont {Rogge}, \citenamefont {Salfi},\ and\ \citenamefont
  {Culcer}}]{Wang2021}%
  \BibitemOpen
  \bibfield  {author} {\bibinfo {author} {\bibfnamefont {Z.}~\bibnamefont
  {Wang}}, \bibinfo {author} {\bibfnamefont {E.}~\bibnamefont {Marcellina}},
  \bibinfo {author} {\bibfnamefont {A.}~\bibnamefont {Hamilton}}, \bibinfo
  {author} {\bibfnamefont {J.~H.}\ \bibnamefont {Cullen}}, \bibinfo {author}
  {\bibfnamefont {S.}~\bibnamefont {Rogge}}, \bibinfo {author} {\bibfnamefont
  {J.}~\bibnamefont {Salfi}},\ and\ \bibinfo {author} {\bibfnamefont
  {D.}~\bibnamefont {Culcer}},\ }\bibfield  {title} {\bibinfo {title} {Optimal
  operation points for ultrafast, highly coherent {Ge} hole spin-orbit
  qubits},\ }\href {https://doi.org/10.1038/s41534-021-00386-2} {\bibfield
  {journal} {\bibinfo  {journal} {npj Quantum Inf}\ }\textbf {\bibinfo {volume}
  {7}},\ \bibinfo {pages} {54} (\bibinfo {year} {2021})}\BibitemShut {NoStop}%
\bibitem [{\citenamefont {Mutter}\ and\ \citenamefont
  {Burkard}(2020{\natexlab{a}})}]{Mutter2020cavitycontrol}%
  \BibitemOpen
  \bibfield  {author} {\bibinfo {author} {\bibfnamefont {P.~M.}\ \bibnamefont
  {Mutter}}\ and\ \bibinfo {author} {\bibfnamefont {G.}~\bibnamefont
  {Burkard}},\ }\bibfield  {title} {\bibinfo {title} {Cavity control over
  heavy-hole spin qubits in inversion-symmetric crystals},\ }\href
  {https://doi.org/10.1103/PhysRevB.102.205412} {\bibfield  {journal} {\bibinfo
   {journal} {Phys. Rev. B}\ }\textbf {\bibinfo {volume} {102}},\ \bibinfo
  {pages} {205412} (\bibinfo {year} {2020}{\natexlab{a}})}\BibitemShut
  {NoStop}%
\bibitem [{\citenamefont {Mutter}\ and\ \citenamefont
  {Burkard}(2021{\natexlab{a}})}]{Mutter2021natural}%
  \BibitemOpen
  \bibfield  {author} {\bibinfo {author} {\bibfnamefont {P.~M.}\ \bibnamefont
  {Mutter}}\ and\ \bibinfo {author} {\bibfnamefont {G.}~\bibnamefont
  {Burkard}},\ }\bibfield  {title} {\bibinfo {title} {Natural heavy-hole
  flopping mode qubit in germanium},\ }\href
  {https://doi.org/10.1103/PhysRevResearch.3.013194} {\bibfield  {journal}
  {\bibinfo  {journal} {Phys. Rev. Research}\ }\textbf {\bibinfo {volume}
  {3}},\ \bibinfo {pages} {013194} (\bibinfo {year}
  {2021}{\natexlab{a}})}\BibitemShut {NoStop}%
\bibitem [{\citenamefont {Benito}\ \emph {et~al.}(2017)\citenamefont {Benito},
  \citenamefont {Mi}, \citenamefont {Taylor}, \citenamefont {Petta},\ and\
  \citenamefont {Burkard}}]{Benito2017}%
  \BibitemOpen
  \bibfield  {author} {\bibinfo {author} {\bibfnamefont {M.}~\bibnamefont
  {Benito}}, \bibinfo {author} {\bibfnamefont {X.}~\bibnamefont {Mi}}, \bibinfo
  {author} {\bibfnamefont {J.~M.}\ \bibnamefont {Taylor}}, \bibinfo {author}
  {\bibfnamefont {J.~R.}\ \bibnamefont {Petta}},\ and\ \bibinfo {author}
  {\bibfnamefont {G.}~\bibnamefont {Burkard}},\ }\bibfield  {title} {\bibinfo
  {title} {Input-output theory for spin-photon coupling in {Si} double quantum
  dots},\ }\href {https://doi.org/10.1103/PhysRevB.96.235434} {\bibfield
  {journal} {\bibinfo  {journal} {Phys. Rev. B}\ }\textbf {\bibinfo {volume}
  {96}},\ \bibinfo {pages} {235434} (\bibinfo {year} {2017})}\BibitemShut
  {NoStop}%
\bibitem [{\citenamefont {Mi}\ \emph {et~al.}(2018)\citenamefont {Mi},
  \citenamefont {Benito}, \citenamefont {Putz}, \citenamefont {Zajac},
  \citenamefont {Taylor}, \citenamefont {Burkard},\ and\ \citenamefont
  {Petta}}]{Mi2018}%
  \BibitemOpen
  \bibfield  {author} {\bibinfo {author} {\bibfnamefont {X.}~\bibnamefont
  {Mi}}, \bibinfo {author} {\bibfnamefont {M.}~\bibnamefont {Benito}}, \bibinfo
  {author} {\bibfnamefont {S.}~\bibnamefont {Putz}}, \bibinfo {author}
  {\bibfnamefont {D.~M.}\ \bibnamefont {Zajac}}, \bibinfo {author}
  {\bibfnamefont {J.~M.}\ \bibnamefont {Taylor}}, \bibinfo {author}
  {\bibfnamefont {G.}~\bibnamefont {Burkard}},\ and\ \bibinfo {author}
  {\bibfnamefont {J.~R.}\ \bibnamefont {Petta}},\ }\bibfield  {title} {\bibinfo
  {title} {A coherent spin-photon interface in silicon},\ }\href
  {https://doi.org/10.1038/nature25769} {\bibfield  {journal} {\bibinfo
  {journal} {Nature}\ }\textbf {\bibinfo {volume} {555}},\ \bibinfo {pages}
  {599} (\bibinfo {year} {2018})}\BibitemShut {NoStop}%
\bibitem [{\citenamefont {Benito}\ \emph
  {et~al.}(2019{\natexlab{a}})\citenamefont {Benito}, \citenamefont {Petta},\
  and\ \citenamefont {Burkard}}]{Benito2019a}%
  \BibitemOpen
  \bibfield  {author} {\bibinfo {author} {\bibfnamefont {M.}~\bibnamefont
  {Benito}}, \bibinfo {author} {\bibfnamefont {J.~R.}\ \bibnamefont {Petta}},\
  and\ \bibinfo {author} {\bibfnamefont {G.}~\bibnamefont {Burkard}},\
  }\bibfield  {title} {\bibinfo {title} {Optimized cavity-mediated dispersive
  two-qubit gates between spin qubits},\ }\href
  {https://doi.org/10.1103/PhysRevB.100.081412} {\bibfield  {journal} {\bibinfo
   {journal} {Phys. Rev. B}\ }\textbf {\bibinfo {volume} {100}},\ \bibinfo
  {pages} {081412} (\bibinfo {year} {2019}{\natexlab{a}})}\BibitemShut
  {NoStop}%
\bibitem [{\citenamefont {Benito}\ \emph
  {et~al.}(2019{\natexlab{b}})\citenamefont {Benito}, \citenamefont {Croot},
  \citenamefont {Adelsberger}, \citenamefont {Putz}, \citenamefont {Mi},
  \citenamefont {Petta},\ and\ \citenamefont {Burkard}}]{Benito2019b}%
  \BibitemOpen
  \bibfield  {author} {\bibinfo {author} {\bibfnamefont {M.}~\bibnamefont
  {Benito}}, \bibinfo {author} {\bibfnamefont {X.}~\bibnamefont {Croot}},
  \bibinfo {author} {\bibfnamefont {C.}~\bibnamefont {Adelsberger}}, \bibinfo
  {author} {\bibfnamefont {S.}~\bibnamefont {Putz}}, \bibinfo {author}
  {\bibfnamefont {X.}~\bibnamefont {Mi}}, \bibinfo {author} {\bibfnamefont
  {J.~R.}\ \bibnamefont {Petta}},\ and\ \bibinfo {author} {\bibfnamefont
  {G.}~\bibnamefont {Burkard}},\ }\bibfield  {title} {\bibinfo {title}
  {Electric-field control and noise protection of the flopping-mode spin
  qubit},\ }\href {https://doi.org/10.1103/PhysRevB.100.125430} {\bibfield
  {journal} {\bibinfo  {journal} {Phys. Rev. B}\ }\textbf {\bibinfo {volume}
  {100}},\ \bibinfo {pages} {125430} (\bibinfo {year}
  {2019}{\natexlab{b}})}\BibitemShut {NoStop}%
\bibitem [{\citenamefont {Croot}\ \emph {et~al.}(2020)\citenamefont {Croot},
  \citenamefont {Mi}, \citenamefont {Putz}, \citenamefont {Benito},
  \citenamefont {Borjans}, \citenamefont {Burkard},\ and\ \citenamefont
  {Petta}}]{Croot2020}%
  \BibitemOpen
  \bibfield  {author} {\bibinfo {author} {\bibfnamefont {X.}~\bibnamefont
  {Croot}}, \bibinfo {author} {\bibfnamefont {X.}~\bibnamefont {Mi}}, \bibinfo
  {author} {\bibfnamefont {S.}~\bibnamefont {Putz}}, \bibinfo {author}
  {\bibfnamefont {M.}~\bibnamefont {Benito}}, \bibinfo {author} {\bibfnamefont
  {F.}~\bibnamefont {Borjans}}, \bibinfo {author} {\bibfnamefont
  {G.}~\bibnamefont {Burkard}},\ and\ \bibinfo {author} {\bibfnamefont {J.~R.}\
  \bibnamefont {Petta}},\ }\bibfield  {title} {\bibinfo {title} {Flopping-mode
  electric dipole spin resonance},\ }\href
  {https://doi.org/10.1103/PhysRevResearch.2.012006} {\bibfield  {journal}
  {\bibinfo  {journal} {Phys. Rev. Research}\ }\textbf {\bibinfo {volume}
  {2}},\ \bibinfo {pages} {012006} (\bibinfo {year} {2020})}\BibitemShut
  {NoStop}%
\bibitem [{\citenamefont {Levy}(2002)}]{Levy2002}%
  \BibitemOpen
  \bibfield  {author} {\bibinfo {author} {\bibfnamefont {J.}~\bibnamefont
  {Levy}},\ }\bibfield  {title} {\bibinfo {title} {Universal quantum
  computation with spin-$1/2$ pairs and heisenberg exchange},\ }\href
  {https://doi.org/10.1103/PhysRevLett.89.147902} {\bibfield  {journal}
  {\bibinfo  {journal} {Phys. Rev. Lett.}\ }\textbf {\bibinfo {volume} {89}},\
  \bibinfo {pages} {147902} (\bibinfo {year} {2002})}\BibitemShut {NoStop}%
\bibitem [{\citenamefont {Petta}\ \emph {et~al.}(2005)\citenamefont {Petta},
  \citenamefont {Johnson}, \citenamefont {Taylor}, \citenamefont {Laird},
  \citenamefont {Yacoby}, \citenamefont {Lukin}, \citenamefont {Marcus},
  \citenamefont {Hanson},\ and\ \citenamefont {Gossard}}]{Petta2005}%
  \BibitemOpen
  \bibfield  {author} {\bibinfo {author} {\bibfnamefont {J.~R.}\ \bibnamefont
  {Petta}}, \bibinfo {author} {\bibfnamefont {A.~C.}\ \bibnamefont {Johnson}},
  \bibinfo {author} {\bibfnamefont {J.~M.}\ \bibnamefont {Taylor}}, \bibinfo
  {author} {\bibfnamefont {E.~A.}\ \bibnamefont {Laird}}, \bibinfo {author}
  {\bibfnamefont {A.}~\bibnamefont {Yacoby}}, \bibinfo {author} {\bibfnamefont
  {M.~D.}\ \bibnamefont {Lukin}}, \bibinfo {author} {\bibfnamefont {C.~M.}\
  \bibnamefont {Marcus}}, \bibinfo {author} {\bibfnamefont {M.~P.}\
  \bibnamefont {Hanson}},\ and\ \bibinfo {author} {\bibfnamefont {A.~C.}\
  \bibnamefont {Gossard}},\ }\bibfield  {title} {\bibinfo {title} {Coherent
  manipulation of coupled electron spins in semiconductor quantum dots},\
  }\href {https://doi.org/10.1126/science.1116955} {\bibfield  {journal}
  {\bibinfo  {journal} {Science}\ }\textbf {\bibinfo {volume} {309}},\ \bibinfo
  {pages} {2180} (\bibinfo {year} {2005})}\BibitemShut {NoStop}%
\bibitem [{\citenamefont {Foletti}\ \emph {et~al.}(2009)\citenamefont
  {Foletti}, \citenamefont {Bluhm}, \citenamefont {Mahalu}, \citenamefont
  {Umansky},\ and\ \citenamefont {Yacoby}}]{Foletti2009}%
  \BibitemOpen
  \bibfield  {author} {\bibinfo {author} {\bibfnamefont {S.}~\bibnamefont
  {Foletti}}, \bibinfo {author} {\bibfnamefont {H.}~\bibnamefont {Bluhm}},
  \bibinfo {author} {\bibfnamefont {D.}~\bibnamefont {Mahalu}}, \bibinfo
  {author} {\bibfnamefont {V.}~\bibnamefont {Umansky}},\ and\ \bibinfo {author}
  {\bibfnamefont {A.}~\bibnamefont {Yacoby}},\ }\bibfield  {title} {\bibinfo
  {title} {Universal quantum control of two-electron spin quantum bits using
  dynamic nuclear polarization},\ }\href {https://doi.org/10.1038/nphys1424}
  {\bibfield  {journal} {\bibinfo  {journal} {Nature Phys}\ }\textbf {\bibinfo
  {volume} {5}},\ \bibinfo {pages} {903} (\bibinfo {year} {2009})}\BibitemShut
  {NoStop}%
\bibitem [{\citenamefont {Wu}\ \emph {et~al.}(2014)\citenamefont {Wu},
  \citenamefont {Ward}, \citenamefont {Prance}, \citenamefont {Kim},
  \citenamefont {Gamble}, \citenamefont {Mohr}, \citenamefont {Shi},
  \citenamefont {Savage}, \citenamefont {Lagally}, \citenamefont {Friesen},
  \citenamefont {Coppersmith},\ and\ \citenamefont {Eriksson}}]{Wu2014}%
  \BibitemOpen
  \bibfield  {author} {\bibinfo {author} {\bibfnamefont {X.}~\bibnamefont
  {Wu}}, \bibinfo {author} {\bibfnamefont {D.~R.}\ \bibnamefont {Ward}},
  \bibinfo {author} {\bibfnamefont {J.~R.}\ \bibnamefont {Prance}}, \bibinfo
  {author} {\bibfnamefont {D.}~\bibnamefont {Kim}}, \bibinfo {author}
  {\bibfnamefont {J.~K.}\ \bibnamefont {Gamble}}, \bibinfo {author}
  {\bibfnamefont {R.~T.}\ \bibnamefont {Mohr}}, \bibinfo {author}
  {\bibfnamefont {Z.}~\bibnamefont {Shi}}, \bibinfo {author} {\bibfnamefont
  {D.~E.}\ \bibnamefont {Savage}}, \bibinfo {author} {\bibfnamefont {M.~G.}\
  \bibnamefont {Lagally}}, \bibinfo {author} {\bibfnamefont {M.}~\bibnamefont
  {Friesen}}, \bibinfo {author} {\bibfnamefont {S.~N.}\ \bibnamefont
  {Coppersmith}},\ and\ \bibinfo {author} {\bibfnamefont {M.~A.}\ \bibnamefont
  {Eriksson}},\ }\bibfield  {title} {\bibinfo {title} {Two-axis control of a
  singlet{\textendash}triplet qubit with an integrated micromagnet},\ }\href
  {https://doi.org/10.1073/pnas.1412230111} {\bibfield  {journal} {\bibinfo
  {journal} {Proceedings of the National Academy of Sciences}\ }\textbf
  {\bibinfo {volume} {111}},\ \bibinfo {pages} {11938} (\bibinfo {year}
  {2014})}\BibitemShut {NoStop}%
\bibitem [{\citenamefont {Nichol}\ \emph {et~al.}(2017)\citenamefont {Nichol},
  \citenamefont {Orona}, \citenamefont {Harvey}, \citenamefont {Fallahi},
  \citenamefont {Gardner}, \citenamefont {Manfra},\ and\ \citenamefont
  {Yacoby}}]{Nichol2017}%
  \BibitemOpen
  \bibfield  {author} {\bibinfo {author} {\bibfnamefont {J.~M.}\ \bibnamefont
  {Nichol}}, \bibinfo {author} {\bibfnamefont {L.~A.}\ \bibnamefont {Orona}},
  \bibinfo {author} {\bibfnamefont {S.~P.}\ \bibnamefont {Harvey}}, \bibinfo
  {author} {\bibfnamefont {S.}~\bibnamefont {Fallahi}}, \bibinfo {author}
  {\bibfnamefont {G.~C.}\ \bibnamefont {Gardner}}, \bibinfo {author}
  {\bibfnamefont {M.~J.}\ \bibnamefont {Manfra}},\ and\ \bibinfo {author}
  {\bibfnamefont {A.}~\bibnamefont {Yacoby}},\ }\bibfield  {title} {\bibinfo
  {title} {High-fidelity entangling gate for double-quantum-dot spin qubits},\
  }\href {https://doi.org/10.1038/s41534-016-0003-1} {\bibfield  {journal}
  {\bibinfo  {journal} {npj Quantum Inf}\ }\textbf {\bibinfo {volume} {3}},\
  \bibinfo {pages} {3} (\bibinfo {year} {2017})}\BibitemShut {NoStop}%
\bibitem [{\citenamefont {Koppens}\ \emph {et~al.}(2005)\citenamefont
  {Koppens}, \citenamefont {Folk}, \citenamefont {Elzerman}, \citenamefont
  {Hanson}, \citenamefont {van Beveren}, \citenamefont {Vink}, \citenamefont
  {Tranitz}, \citenamefont {Wegscheider}, \citenamefont {Kouwenhoven},\ and\
  \citenamefont {Vandersypen}}]{Koppens2005}%
  \BibitemOpen
  \bibfield  {author} {\bibinfo {author} {\bibfnamefont {F.~H.~L.}\
  \bibnamefont {Koppens}}, \bibinfo {author} {\bibfnamefont {J.~A.}\
  \bibnamefont {Folk}}, \bibinfo {author} {\bibfnamefont {J.~M.}\ \bibnamefont
  {Elzerman}}, \bibinfo {author} {\bibfnamefont {R.}~\bibnamefont {Hanson}},
  \bibinfo {author} {\bibfnamefont {L.~H.~W.}\ \bibnamefont {van Beveren}},
  \bibinfo {author} {\bibfnamefont {I.~T.}\ \bibnamefont {Vink}}, \bibinfo
  {author} {\bibfnamefont {H.~P.}\ \bibnamefont {Tranitz}}, \bibinfo {author}
  {\bibfnamefont {W.}~\bibnamefont {Wegscheider}}, \bibinfo {author}
  {\bibfnamefont {L.~P.}\ \bibnamefont {Kouwenhoven}},\ and\ \bibinfo {author}
  {\bibfnamefont {L.~M.~K.}\ \bibnamefont {Vandersypen}},\ }\bibfield  {title}
  {\bibinfo {title} {Control and detection of singlet-triplet mixing in a
  random nuclear field},\ }\href {https://doi.org/10.1126/science.1113719}
  {\bibfield  {journal} {\bibinfo  {journal} {Science}\ }\textbf {\bibinfo
  {volume} {309}},\ \bibinfo {pages} {1346} (\bibinfo {year}
  {2005})}\BibitemShut {NoStop}%
\bibitem [{\citenamefont {Maune}\ \emph {et~al.}(2014)\citenamefont {Maune},
  \citenamefont {Borselli}, \citenamefont {Huang}, \citenamefont {Ladd},
  \citenamefont {Deelman}, \citenamefont {Holabird}, \citenamefont {Kiselev},
  \citenamefont {Alvarado-Rodriguez}, \citenamefont {Ross}, \citenamefont
  {Schmitz}, \citenamefont {Sokolich}, \citenamefont {Watson}, \citenamefont
  {Gyure},\ and\ \citenamefont {Hunter}}]{Maune2012}%
  \BibitemOpen
  \bibfield  {author} {\bibinfo {author} {\bibfnamefont {B.~M.}\ \bibnamefont
  {Maune}}, \bibinfo {author} {\bibfnamefont {M.~G.}\ \bibnamefont {Borselli}},
  \bibinfo {author} {\bibfnamefont {B.}~\bibnamefont {Huang}}, \bibinfo
  {author} {\bibfnamefont {T.~D.}\ \bibnamefont {Ladd}}, \bibinfo {author}
  {\bibfnamefont {P.~W.}\ \bibnamefont {Deelman}}, \bibinfo {author}
  {\bibfnamefont {K.~S.}\ \bibnamefont {Holabird}}, \bibinfo {author}
  {\bibfnamefont {A.~A.}\ \bibnamefont {Kiselev}}, \bibinfo {author}
  {\bibfnamefont {I.}~\bibnamefont {Alvarado-Rodriguez}}, \bibinfo {author}
  {\bibfnamefont {R.~S.}\ \bibnamefont {Ross}}, \bibinfo {author}
  {\bibfnamefont {A.~E.}\ \bibnamefont {Schmitz}}, \bibinfo {author}
  {\bibfnamefont {M.}~\bibnamefont {Sokolich}}, \bibinfo {author}
  {\bibfnamefont {C.~A.}\ \bibnamefont {Watson}}, \bibinfo {author}
  {\bibfnamefont {M.~F.}\ \bibnamefont {Gyure}},\ and\ \bibinfo {author}
  {\bibfnamefont {A.~T.}\ \bibnamefont {Hunter}},\ }\bibfield  {title}
  {\bibinfo {title} {Coherent singlet-triplet oscillations in a silicon-based
  double quantum dot},\ }\href {https://doi.org/10.1038/nature10707} {\bibfield
   {journal} {\bibinfo  {journal} {Nature}\ }\textbf {\bibinfo {volume}
  {481}},\ \bibinfo {pages} {344} (\bibinfo {year} {2014})}\BibitemShut
  {NoStop}%
\bibitem [{\citenamefont {Jock}\ \emph {et~al.}(2018)\citenamefont {Jock},
  \citenamefont {Jacobson}, \citenamefont {Harvey-Collard}, \citenamefont
  {Mounce}, \citenamefont {Srinivasa}, \citenamefont {Ward}, \citenamefont
  {Anderson}, \citenamefont {Manginell}, \citenamefont {Wendt}, \citenamefont
  {Rudolph}, \citenamefont {Pluym}, \citenamefont {Gamble}, \citenamefont
  {Baczewski}, \citenamefont {Witzel},\ and\ \citenamefont
  {Carroll}}]{Jock2018}%
  \BibitemOpen
  \bibfield  {author} {\bibinfo {author} {\bibfnamefont {R.~M.}\ \bibnamefont
  {Jock}}, \bibinfo {author} {\bibfnamefont {N.~T.}\ \bibnamefont {Jacobson}},
  \bibinfo {author} {\bibfnamefont {P.}~\bibnamefont {Harvey-Collard}},
  \bibinfo {author} {\bibfnamefont {A.~M.}\ \bibnamefont {Mounce}}, \bibinfo
  {author} {\bibfnamefont {V.}~\bibnamefont {Srinivasa}}, \bibinfo {author}
  {\bibfnamefont {D.~R.}\ \bibnamefont {Ward}}, \bibinfo {author}
  {\bibfnamefont {J.}~\bibnamefont {Anderson}}, \bibinfo {author}
  {\bibfnamefont {R.}~\bibnamefont {Manginell}}, \bibinfo {author}
  {\bibfnamefont {J.~R.}\ \bibnamefont {Wendt}}, \bibinfo {author}
  {\bibfnamefont {M.}~\bibnamefont {Rudolph}}, \bibinfo {author} {\bibfnamefont
  {T.}~\bibnamefont {Pluym}}, \bibinfo {author} {\bibfnamefont {J.~K.}\
  \bibnamefont {Gamble}}, \bibinfo {author} {\bibfnamefont {A.~D.}\
  \bibnamefont {Baczewski}}, \bibinfo {author} {\bibfnamefont {W.~M.}\
  \bibnamefont {Witzel}},\ and\ \bibinfo {author} {\bibfnamefont {M.~S.}\
  \bibnamefont {Carroll}},\ }\bibfield  {title} {\bibinfo {title} {A silicon
  metal-oxide-semiconductor electron spin-orbit qubit},\ }\href
  {https://doi.org/10.1038/s41467-018-04200-0} {\bibfield  {journal} {\bibinfo
  {journal} {Nat Commun}\ }\textbf {\bibinfo {volume} {9}},\ \bibinfo {pages}
  {1768} (\bibinfo {year} {2018})}\BibitemShut {NoStop}%
\bibitem [{\citenamefont {Hofmann}\ \emph {et~al.}()\citenamefont {Hofmann},
  \citenamefont {Jirovec}, \citenamefont {Borovkov}, \citenamefont {Prieto},
  \citenamefont {Ballabio}, \citenamefont {Frigerio}, \citenamefont
  {Chrastina}, \citenamefont {Isella},\ and\ \citenamefont
  {Katsaros}}]{Hofmann2019arXiv}%
  \BibitemOpen
  \bibfield  {author} {\bibinfo {author} {\bibfnamefont {A.}~\bibnamefont
  {Hofmann}}, \bibinfo {author} {\bibfnamefont {D.}~\bibnamefont {Jirovec}},
  \bibinfo {author} {\bibfnamefont {M.}~\bibnamefont {Borovkov}}, \bibinfo
  {author} {\bibfnamefont {I.}~\bibnamefont {Prieto}}, \bibinfo {author}
  {\bibfnamefont {A.}~\bibnamefont {Ballabio}}, \bibinfo {author}
  {\bibfnamefont {J.}~\bibnamefont {Frigerio}}, \bibinfo {author}
  {\bibfnamefont {D.}~\bibnamefont {Chrastina}}, \bibinfo {author}
  {\bibfnamefont {G.}~\bibnamefont {Isella}},\ and\ \bibinfo {author}
  {\bibfnamefont {G.}~\bibnamefont {Katsaros}},\ }\href@noop {} {}\Eprint
  {https://arxiv.org/abs/1910.05841} {arXiv:1910.05841 [cond-mat.mes-hall]}
  \BibitemShut {NoStop}%
\bibitem [{\citenamefont {Jirovec}\ \emph {et~al.}(2021)\citenamefont
  {Jirovec}, \citenamefont {Hofmann}, \citenamefont {Ballabio}, \citenamefont
  {Mutter}, \citenamefont {Tavani}, \citenamefont {Botifoll}, \citenamefont
  {Crippa}, \citenamefont {Kukucka}, \citenamefont {Sagi}, \citenamefont
  {Martins}, \citenamefont {Saez-Mollejo}, \citenamefont {Prieto},
  \citenamefont {Borovkov}, \citenamefont {Arbiol}, \citenamefont {Chrastina},
  \citenamefont {Isella},\ and\ \citenamefont {Katsaros}}]{Jirovec2021}%
  \BibitemOpen
  \bibfield  {author} {\bibinfo {author} {\bibfnamefont {D.}~\bibnamefont
  {Jirovec}}, \bibinfo {author} {\bibfnamefont {A.}~\bibnamefont {Hofmann}},
  \bibinfo {author} {\bibfnamefont {A.}~\bibnamefont {Ballabio}}, \bibinfo
  {author} {\bibfnamefont {P.~M.}\ \bibnamefont {Mutter}}, \bibinfo {author}
  {\bibfnamefont {G.}~\bibnamefont {Tavani}}, \bibinfo {author} {\bibfnamefont
  {M.}~\bibnamefont {Botifoll}}, \bibinfo {author} {\bibfnamefont
  {A.}~\bibnamefont {Crippa}}, \bibinfo {author} {\bibfnamefont
  {J.}~\bibnamefont {Kukucka}}, \bibinfo {author} {\bibfnamefont
  {O.}~\bibnamefont {Sagi}}, \bibinfo {author} {\bibfnamefont {F.}~\bibnamefont
  {Martins}}, \bibinfo {author} {\bibfnamefont {J.}~\bibnamefont
  {Saez-Mollejo}}, \bibinfo {author} {\bibfnamefont {I.}~\bibnamefont
  {Prieto}}, \bibinfo {author} {\bibfnamefont {M.}~\bibnamefont {Borovkov}},
  \bibinfo {author} {\bibfnamefont {J.}~\bibnamefont {Arbiol}}, \bibinfo
  {author} {\bibfnamefont {D.}~\bibnamefont {Chrastina}}, \bibinfo {author}
  {\bibfnamefont {G.}~\bibnamefont {Isella}},\ and\ \bibinfo {author}
  {\bibfnamefont {G.}~\bibnamefont {Katsaros}},\ }\bibfield  {title} {\bibinfo
  {title} {A singlet-triplet hole spin qubit in planar {Ge}},\ }\href
  {https://doi.org/10.1038/s41563-021-01022-2} {\bibfield  {journal} {\bibinfo
  {journal} {Nature Materials}\ } (\bibinfo {year} {2021})}\BibitemShut
  {NoStop}%
\bibitem [{\citenamefont {Burkard}\ \emph {et~al.}(2020)\citenamefont
  {Burkard}, \citenamefont {Gullans}, \citenamefont {Mi},\ and\ \citenamefont
  {Petta}}]{Burkard2020}%
  \BibitemOpen
  \bibfield  {author} {\bibinfo {author} {\bibfnamefont {G.}~\bibnamefont
  {Burkard}}, \bibinfo {author} {\bibfnamefont {M.~J.}\ \bibnamefont
  {Gullans}}, \bibinfo {author} {\bibfnamefont {X.}~\bibnamefont {Mi}},\ and\
  \bibinfo {author} {\bibfnamefont {J.}~\bibnamefont {Petta}},\ }\bibfield
  {title} {\bibinfo {title} {Superconductor–semiconductor hybrid-circuit
  quantum electrodynamics},\ }\href {https://doi.org/10.1038/s42254-019-0135-2}
  {\bibfield  {journal} {\bibinfo  {journal} {Nat. Rev. Phys.}\ }\textbf
  {\bibinfo {volume} {2}},\ \bibinfo {pages} {129} (\bibinfo {year}
  {2020})}\BibitemShut {NoStop}%
\bibitem [{\citenamefont {Fischer}\ \emph {et~al.}(2009)\citenamefont
  {Fischer}, \citenamefont {Trif}, \citenamefont {Coish},\ and\ \citenamefont
  {Loss}}]{Fischer2009}%
  \BibitemOpen
  \bibfield  {author} {\bibinfo {author} {\bibfnamefont {J.}~\bibnamefont
  {Fischer}}, \bibinfo {author} {\bibfnamefont {M.}~\bibnamefont {Trif}},
  \bibinfo {author} {\bibfnamefont {W.~A.}\ \bibnamefont {Coish}},\ and\
  \bibinfo {author} {\bibfnamefont {D.}~\bibnamefont {Loss}},\ }\bibfield
  {title} {\bibinfo {title} {Spin interactions, relaxation and decoherence in
  quantum dots},\ }\href
  {https://doi.org/https://doi.org/10.1016/j.ssc.2009.04.033} {\bibfield
  {journal} {\bibinfo  {journal} {Solid State Communications}\ }\textbf
  {\bibinfo {volume} {149}},\ \bibinfo {pages} {1443} (\bibinfo {year}
  {2009})}\BibitemShut {NoStop}%
\bibitem [{\citenamefont {Fischer}\ and\ \citenamefont
  {Loss}(2010)}]{Fischer2010}%
  \BibitemOpen
  \bibfield  {author} {\bibinfo {author} {\bibfnamefont {J.}~\bibnamefont
  {Fischer}}\ and\ \bibinfo {author} {\bibfnamefont {D.}~\bibnamefont {Loss}},\
  }\bibfield  {title} {\bibinfo {title} {Hybridization and spin decoherence in
  heavy-hole quantum dots},\ }\href
  {https://doi.org/10.1103/PhysRevLett.105.266603} {\bibfield  {journal}
  {\bibinfo  {journal} {Phys. Rev. Lett.}\ }\textbf {\bibinfo {volume} {105}},\
  \bibinfo {pages} {266603} (\bibinfo {year} {2010})}\BibitemShut {NoStop}%
\bibitem [{\citenamefont {Maier}\ and\ \citenamefont {Loss}(2012)}]{Maier2012}%
  \BibitemOpen
  \bibfield  {author} {\bibinfo {author} {\bibfnamefont {F.}~\bibnamefont
  {Maier}}\ and\ \bibinfo {author} {\bibfnamefont {D.}~\bibnamefont {Loss}},\
  }\bibfield  {title} {\bibinfo {title} {Effect of strain on hyperfine-induced
  hole-spin decoherence in quantum dots},\ }\href
  {https://doi.org/10.1103/PhysRevB.85.195323} {\bibfield  {journal} {\bibinfo
  {journal} {Phys. Rev. B}\ }\textbf {\bibinfo {volume} {85}},\ \bibinfo
  {pages} {195323} (\bibinfo {year} {2012})}\BibitemShut {NoStop}%
\bibitem [{\citenamefont {Sigillito}\ \emph {et~al.}(2015)\citenamefont
  {Sigillito}, \citenamefont {Jock}, \citenamefont {Tyryshkin}, \citenamefont
  {Beeman}, \citenamefont {Haller}, \citenamefont {Itoh},\ and\ \citenamefont
  {Lyon}}]{Sigillito2015}%
  \BibitemOpen
  \bibfield  {author} {\bibinfo {author} {\bibfnamefont {A.~J.}\ \bibnamefont
  {Sigillito}}, \bibinfo {author} {\bibfnamefont {R.~M.}\ \bibnamefont {Jock}},
  \bibinfo {author} {\bibfnamefont {A.~M.}\ \bibnamefont {Tyryshkin}}, \bibinfo
  {author} {\bibfnamefont {J.~W.}\ \bibnamefont {Beeman}}, \bibinfo {author}
  {\bibfnamefont {E.~E.}\ \bibnamefont {Haller}}, \bibinfo {author}
  {\bibfnamefont {K.~M.}\ \bibnamefont {Itoh}},\ and\ \bibinfo {author}
  {\bibfnamefont {S.~A.}\ \bibnamefont {Lyon}},\ }\bibfield  {title} {\bibinfo
  {title} {Electron spin coherence of shallow donors in natural and
  isotopically enriched germanium},\ }\href
  {https://doi.org/10.1103/PhysRevLett.115.247601} {\bibfield  {journal}
  {\bibinfo  {journal} {Phys. Rev. Lett.}\ }\textbf {\bibinfo {volume} {115}},\
  \bibinfo {pages} {247601} (\bibinfo {year} {2015})}\BibitemShut {NoStop}%
\bibitem [{\citenamefont {Scappucci}\ \emph {et~al.}(2020)\citenamefont
  {Scappucci}, \citenamefont {Kloeffel}, \citenamefont {Zwanenburg},
  \citenamefont {Loss}, \citenamefont {Myronov}, \citenamefont {Zhang},
  \citenamefont {Franceschi}, \citenamefont {Katsaros},\ and\ \citenamefont
  {Veldhorst}}]{Scappucci2020review}%
  \BibitemOpen
  \bibfield  {author} {\bibinfo {author} {\bibfnamefont {G.}~\bibnamefont
  {Scappucci}}, \bibinfo {author} {\bibfnamefont {C.}~\bibnamefont {Kloeffel}},
  \bibinfo {author} {\bibfnamefont {F.~A.}\ \bibnamefont {Zwanenburg}},
  \bibinfo {author} {\bibfnamefont {D.}~\bibnamefont {Loss}}, \bibinfo {author}
  {\bibfnamefont {M.}~\bibnamefont {Myronov}}, \bibinfo {author} {\bibfnamefont
  {J.-J.}\ \bibnamefont {Zhang}}, \bibinfo {author} {\bibfnamefont {S.~D.}\
  \bibnamefont {Franceschi}}, \bibinfo {author} {\bibfnamefont
  {G.}~\bibnamefont {Katsaros}},\ and\ \bibinfo {author} {\bibfnamefont
  {M.}~\bibnamefont {Veldhorst}},\ }\bibfield  {title} {\bibinfo {title} {The
  germanium quantum information route},\ }\href
  {https://doi.org/10.1038/s41578-020-00262-z} {\bibfield  {journal} {\bibinfo
  {journal} {Nat. Rev. Mater.}\ } (\bibinfo {year} {2020})}\BibitemShut
  {NoStop}%
\bibitem [{\citenamefont {Bulaev}\ and\ \citenamefont
  {Loss}(2005{\natexlab{a}})}]{Bulaev2005}%
  \BibitemOpen
  \bibfield  {author} {\bibinfo {author} {\bibfnamefont {D.~V.}\ \bibnamefont
  {Bulaev}}\ and\ \bibinfo {author} {\bibfnamefont {D.}~\bibnamefont {Loss}},\
  }\bibfield  {title} {\bibinfo {title} {Spin relaxation and decoherence of
  holes in quantum dots},\ }\href
  {https://doi.org/10.1103/PhysRevLett.95.076805} {\bibfield  {journal}
  {\bibinfo  {journal} {Phys. Rev. Lett.}\ }\textbf {\bibinfo {volume} {95}},\
  \bibinfo {pages} {076805} (\bibinfo {year} {2005}{\natexlab{a}})}\BibitemShut
  {NoStop}%
\bibitem [{\citenamefont {Bulaev}\ and\ \citenamefont
  {Loss}(2005{\natexlab{b}})}]{Bulaev2005b}%
  \BibitemOpen
  \bibfield  {author} {\bibinfo {author} {\bibfnamefont {D.~V.}\ \bibnamefont
  {Bulaev}}\ and\ \bibinfo {author} {\bibfnamefont {D.}~\bibnamefont {Loss}},\
  }\bibfield  {title} {\bibinfo {title} {Spin relaxation and anticrossing in
  quantum dots: Rashba versus dresselhaus spin-orbit coupling},\ }\href
  {https://doi.org/10.1103/PhysRevB.71.205324} {\bibfield  {journal} {\bibinfo
  {journal} {Phys. Rev. B}\ }\textbf {\bibinfo {volume} {71}},\ \bibinfo
  {pages} {205324} (\bibinfo {year} {2005}{\natexlab{b}})}\BibitemShut
  {NoStop}%
\bibitem [{\citenamefont {Bulaev}\ and\ \citenamefont
  {Loss}(2007)}]{Bulaev2007}%
  \BibitemOpen
  \bibfield  {author} {\bibinfo {author} {\bibfnamefont {D.~V.}\ \bibnamefont
  {Bulaev}}\ and\ \bibinfo {author} {\bibfnamefont {D.}~\bibnamefont {Loss}},\
  }\bibfield  {title} {\bibinfo {title} {Electric dipole spin resonance for
  heavy holes in quantum dots},\ }\href
  {https://doi.org/10.1103/PhysRevLett.98.097202} {\bibfield  {journal}
  {\bibinfo  {journal} {Phys. Rev. Lett.}\ }\textbf {\bibinfo {volume} {98}},\
  \bibinfo {pages} {097202} (\bibinfo {year} {2007})}\BibitemShut {NoStop}%
\bibitem [{\citenamefont {Barenco}\ \emph {et~al.}(1995)\citenamefont
  {Barenco}, \citenamefont {Bennett}, \citenamefont {Cleve}, \citenamefont
  {DiVincenzo}, \citenamefont {Margolus}, \citenamefont {Shor}, \citenamefont
  {Sleator}, \citenamefont {Smolin},\ and\ \citenamefont
  {Weinfurter}}]{Barenco1995}%
  \BibitemOpen
  \bibfield  {author} {\bibinfo {author} {\bibfnamefont {A.}~\bibnamefont
  {Barenco}}, \bibinfo {author} {\bibfnamefont {C.~H.}\ \bibnamefont
  {Bennett}}, \bibinfo {author} {\bibfnamefont {R.}~\bibnamefont {Cleve}},
  \bibinfo {author} {\bibfnamefont {D.~P.}\ \bibnamefont {DiVincenzo}},
  \bibinfo {author} {\bibfnamefont {N.}~\bibnamefont {Margolus}}, \bibinfo
  {author} {\bibfnamefont {P.}~\bibnamefont {Shor}}, \bibinfo {author}
  {\bibfnamefont {T.}~\bibnamefont {Sleator}}, \bibinfo {author} {\bibfnamefont
  {J.~A.}\ \bibnamefont {Smolin}},\ and\ \bibinfo {author} {\bibfnamefont
  {H.}~\bibnamefont {Weinfurter}},\ }\bibfield  {title} {\bibinfo {title}
  {Elementary gates for quantum computation},\ }\href
  {https://doi.org/10.1103/PhysRevA.52.3457} {\bibfield  {journal} {\bibinfo
  {journal} {Phys. Rev. A}\ }\textbf {\bibinfo {volume} {52}},\ \bibinfo
  {pages} {3457} (\bibinfo {year} {1995})}\BibitemShut {NoStop}%
\bibitem [{\citenamefont {Hanson}\ and\ \citenamefont
  {Burkard}(2007)}]{Hanson2007}%
  \BibitemOpen
  \bibfield  {author} {\bibinfo {author} {\bibfnamefont {R.}~\bibnamefont
  {Hanson}}\ and\ \bibinfo {author} {\bibfnamefont {G.}~\bibnamefont
  {Burkard}},\ }\bibfield  {title} {\bibinfo {title} {Universal set of quantum
  gates for double-dot spin qubits with fixed interdot coupling},\ }\href
  {https://doi.org/10.1103/PhysRevLett.98.050502} {\bibfield  {journal}
  {\bibinfo  {journal} {Phys. Rev. Lett.}\ }\textbf {\bibinfo {volume} {98}},\
  \bibinfo {pages} {050502} (\bibinfo {year} {2007})}\BibitemShut {NoStop}%
\bibitem [{\citenamefont {Jouravlev}\ and\ \citenamefont
  {Nazarov}(2006)}]{Jouravlev2006}%
  \BibitemOpen
  \bibfield  {author} {\bibinfo {author} {\bibfnamefont {O.~N.}\ \bibnamefont
  {Jouravlev}}\ and\ \bibinfo {author} {\bibfnamefont {Y.~V.}\ \bibnamefont
  {Nazarov}},\ }\bibfield  {title} {\bibinfo {title} {Electron transport in a
  double quantum dot governed by a nuclear magnetic field},\ }\href
  {https://doi.org/10.1103/PhysRevLett.96.176804} {\bibfield  {journal}
  {\bibinfo  {journal} {Phys. Rev. Lett.}\ }\textbf {\bibinfo {volume} {96}},\
  \bibinfo {pages} {176804} (\bibinfo {year} {2006})}\BibitemShut {NoStop}%
\bibitem [{\citenamefont {Danon}\ and\ \citenamefont
  {Nazarov}(2009)}]{Danon2009}%
  \BibitemOpen
  \bibfield  {author} {\bibinfo {author} {\bibfnamefont {J.}~\bibnamefont
  {Danon}}\ and\ \bibinfo {author} {\bibfnamefont {Y.~V.}\ \bibnamefont
  {Nazarov}},\ }\bibfield  {title} {\bibinfo {title} {Pauli spin blockade in
  the presence of strong spin-orbit coupling},\ }\href
  {https://doi.org/10.1103/PhysRevB.80.041301} {\bibfield  {journal} {\bibinfo
  {journal} {Phys. Rev. B}\ }\textbf {\bibinfo {volume} {80}},\ \bibinfo
  {pages} {041301} (\bibinfo {year} {2009})}\BibitemShut {NoStop}%
\bibitem [{\citenamefont {Mutter}\ and\ \citenamefont
  {Burkard}(2020{\natexlab{b}})}]{Mutter2020}%
  \BibitemOpen
  \bibfield  {author} {\bibinfo {author} {\bibfnamefont {P.~M.}\ \bibnamefont
  {Mutter}}\ and\ \bibinfo {author} {\bibfnamefont {G.}~\bibnamefont
  {Burkard}},\ }\bibfield  {title} {\bibinfo {title} {g-tensor resonance in
  double quantum dots with site-dependent g-tensors},\ }\href
  {https://doi.org/10.1088/2633-4356/ab9c3a} {\bibfield  {journal} {\bibinfo
  {journal} {Mater. Quantum. Technol.}\ }\textbf {\bibinfo {volume} {1}},\
  \bibinfo {pages} {015003} (\bibinfo {year} {2020}{\natexlab{b}})}\BibitemShut
  {NoStop}%
\bibitem [{\citenamefont {Mutter}\ and\ \citenamefont
  {Burkard}(2021{\natexlab{b}})}]{MutterPSB2021}%
  \BibitemOpen
  \bibfield  {author} {\bibinfo {author} {\bibfnamefont {P.~M.}\ \bibnamefont
  {Mutter}}\ and\ \bibinfo {author} {\bibfnamefont {G.}~\bibnamefont
  {Burkard}},\ }\bibfield  {title} {\bibinfo {title} {Pauli spin blockade with
  site-dependent $g$ tensors and spin-polarized leads},\ }\href
  {https://doi.org/10.1103/PhysRevB.103.245412} {\bibfield  {journal} {\bibinfo
   {journal} {Phys. Rev. B}\ }\textbf {\bibinfo {volume} {103}},\ \bibinfo
  {pages} {245412} (\bibinfo {year} {2021}{\natexlab{b}})}\BibitemShut
  {NoStop}%
\bibitem [{\citenamefont {Golovach}\ \emph {et~al.}(2004)\citenamefont
  {Golovach}, \citenamefont {Khaetskii},\ and\ \citenamefont
  {Loss}}]{Golovach2004}%
  \BibitemOpen
  \bibfield  {author} {\bibinfo {author} {\bibfnamefont {V.~N.}\ \bibnamefont
  {Golovach}}, \bibinfo {author} {\bibfnamefont {A.}~\bibnamefont
  {Khaetskii}},\ and\ \bibinfo {author} {\bibfnamefont {D.}~\bibnamefont
  {Loss}},\ }\bibfield  {title} {\bibinfo {title} {Phonon-induced decay of the
  electron spin in quantum dots},\ }\href
  {https://doi.org/10.1103/PhysRevLett.93.016601} {\bibfield  {journal}
  {\bibinfo  {journal} {Phys. Rev. Lett.}\ }\textbf {\bibinfo {volume} {93}},\
  \bibinfo {pages} {016601} (\bibinfo {year} {2004})}\BibitemShut {NoStop}%
\bibitem [{Note1()}]{Note1}%
  \BibitemOpen
  \bibinfo {note} {In addition, one must rescale the triplet states by minus
  one. However, since the physical states are only representatives of an entire
  ray of states in a Hilbert space, this minus only amounts to an irrelevant
  phase.}\BibitemShut {Stop}%
\bibitem [{\citenamefont {Froning}\ \emph
  {et~al.}(2021{\natexlab{a}})\citenamefont {Froning}, \citenamefont
  {Camenzind}, \citenamefont {van~der Molen}, \citenamefont {Li}, \citenamefont
  {Bakkers}, \citenamefont {Zumb\"{u}hl},\ and\ \citenamefont
  {Braakman}}]{Froning2021}%
  \BibitemOpen
  \bibfield  {author} {\bibinfo {author} {\bibfnamefont {F.~N.~M.}\
  \bibnamefont {Froning}}, \bibinfo {author} {\bibfnamefont {L.~C.}\
  \bibnamefont {Camenzind}}, \bibinfo {author} {\bibfnamefont {O.~A.~H.}\
  \bibnamefont {van~der Molen}}, \bibinfo {author} {\bibfnamefont
  {A.}~\bibnamefont {Li}}, \bibinfo {author} {\bibfnamefont {E.~P. A.~M.}\
  \bibnamefont {Bakkers}}, \bibinfo {author} {\bibfnamefont {D.~M.}\
  \bibnamefont {Zumb\"{u}hl}},\ and\ \bibinfo {author} {\bibfnamefont {F.~R.}\
  \bibnamefont {Braakman}},\ }\bibfield  {title} {\bibinfo {title} {Ultrafast
  hole spin qubit with gate-tunable spin-orbit switch},\ }\href
  {https://doi.org/10.1038/s41565-020-00828-6} {\bibfield  {journal} {\bibinfo
  {journal} {Nat. Nanotechnol.}\ }\textbf {\bibinfo {volume} {16}},\ \bibinfo
  {pages} {308} (\bibinfo {year} {2021}{\natexlab{a}})}\BibitemShut {NoStop}%
\bibitem [{\citenamefont {Petta}\ \emph {et~al.}(2010)\citenamefont {Petta},
  \citenamefont {Lu},\ and\ \citenamefont {Gossard}}]{Petta2010}%
  \BibitemOpen
  \bibfield  {author} {\bibinfo {author} {\bibfnamefont {J.~R.}\ \bibnamefont
  {Petta}}, \bibinfo {author} {\bibfnamefont {H.}~\bibnamefont {Lu}},\ and\
  \bibinfo {author} {\bibfnamefont {A.~C.}\ \bibnamefont {Gossard}},\
  }\bibfield  {title} {\bibinfo {title} {A coherent beam splitter for
  electronic spin states},\ }\href {https://doi.org/10.1126/science.1183628}
  {\bibfield  {journal} {\bibinfo  {journal} {Science}\ }\textbf {\bibinfo
  {volume} {327}},\ \bibinfo {pages} {669} (\bibinfo {year}
  {2010})}\BibitemShut {NoStop}%
\bibitem [{\citenamefont {Frank}\ \emph {et~al.}(2020)\citenamefont {Frank},
  \citenamefont {Scher\"ubl}, \citenamefont {Csonka}, \citenamefont
  {Zar\'and},\ and\ \citenamefont {P\'alyi}}]{Gyorgy2020}%
  \BibitemOpen
  \bibfield  {author} {\bibinfo {author} {\bibfnamefont {G.}~\bibnamefont
  {Frank}}, \bibinfo {author} {\bibfnamefont {Z.}~\bibnamefont {Scher\"ubl}},
  \bibinfo {author} {\bibfnamefont {S.}~\bibnamefont {Csonka}}, \bibinfo
  {author} {\bibfnamefont {G.}~\bibnamefont {Zar\'and}},\ and\ \bibinfo
  {author} {\bibfnamefont {A.}~\bibnamefont {P\'alyi}},\ }\bibfield  {title}
  {\bibinfo {title} {Magnetic degeneracy points in interacting two-spin
  systems: Geometrical patterns, topological charge distributions, and their
  stability},\ }\href {https://doi.org/10.1103/PhysRevB.101.245409} {\bibfield
  {journal} {\bibinfo  {journal} {Phys. Rev. B}\ }\textbf {\bibinfo {volume}
  {101}},\ \bibinfo {pages} {245409} (\bibinfo {year} {2020})}\BibitemShut
  {NoStop}%
\bibitem [{\citenamefont {Froning}\ \emph
  {et~al.}(2021{\natexlab{b}})\citenamefont {Froning}, \citenamefont
  {Ran\ifmmode \check{c}\else \v{c}\fi{}i\ifmmode~\acute{c}\else \'{c}\fi{}},
  \citenamefont {Het\'enyi}, \citenamefont {Bosco}, \citenamefont {Rehmann},
  \citenamefont {Li}, \citenamefont {Bakkers}, \citenamefont {Zwanenburg},
  \citenamefont {Loss}, \citenamefont {Zumb\"uhl},\ and\ \citenamefont
  {Braakman}}]{Froning2020b}%
  \BibitemOpen
  \bibfield  {author} {\bibinfo {author} {\bibfnamefont {F.~N.~M.}\
  \bibnamefont {Froning}}, \bibinfo {author} {\bibfnamefont {M.~J.}\
  \bibnamefont {Ran\ifmmode \check{c}\else \v{c}\fi{}i\ifmmode~\acute{c}\else
  \'{c}\fi{}}}, \bibinfo {author} {\bibfnamefont {B.}~\bibnamefont
  {Het\'enyi}}, \bibinfo {author} {\bibfnamefont {S.}~\bibnamefont {Bosco}},
  \bibinfo {author} {\bibfnamefont {M.~K.}\ \bibnamefont {Rehmann}}, \bibinfo
  {author} {\bibfnamefont {A.}~\bibnamefont {Li}}, \bibinfo {author}
  {\bibfnamefont {E.~P. A.~M.}\ \bibnamefont {Bakkers}}, \bibinfo {author}
  {\bibfnamefont {F.~A.}\ \bibnamefont {Zwanenburg}}, \bibinfo {author}
  {\bibfnamefont {D.}~\bibnamefont {Loss}}, \bibinfo {author} {\bibfnamefont
  {D.~M.}\ \bibnamefont {Zumb\"uhl}},\ and\ \bibinfo {author} {\bibfnamefont
  {F.~R.}\ \bibnamefont {Braakman}},\ }\bibfield  {title} {\bibinfo {title}
  {Strong spin-orbit interaction and $g$-factor renormalization of hole spins
  in {Ge}/{Si} nanowire quantum dots},\ }\href
  {https://doi.org/10.1103/PhysRevResearch.3.013081} {\bibfield  {journal}
  {\bibinfo  {journal} {Phys. Rev. Research}\ }\textbf {\bibinfo {volume}
  {3}},\ \bibinfo {pages} {013081} (\bibinfo {year}
  {2021}{\natexlab{b}})}\BibitemShut {NoStop}%
\bibitem [{\citenamefont {Zhang}\ \emph {et~al.}(2021)\citenamefont {Zhang},
  \citenamefont {Liu}, \citenamefont {Gao}, \citenamefont {Xu}, \citenamefont
  {Wang}, \citenamefont {Zhang}, \citenamefont {Cao}, \citenamefont {Wang},
  \citenamefont {Zhang}, \citenamefont {Hu},\ and\ \citenamefont
  {et~al.}}]{Zhang2021}%
  \BibitemOpen
  \bibfield  {author} {\bibinfo {author} {\bibfnamefont {T.}~\bibnamefont
  {Zhang}}, \bibinfo {author} {\bibfnamefont {H.}~\bibnamefont {Liu}}, \bibinfo
  {author} {\bibfnamefont {F.}~\bibnamefont {Gao}}, \bibinfo {author}
  {\bibfnamefont {G.}~\bibnamefont {Xu}}, \bibinfo {author} {\bibfnamefont
  {K.}~\bibnamefont {Wang}}, \bibinfo {author} {\bibfnamefont {X.}~\bibnamefont
  {Zhang}}, \bibinfo {author} {\bibfnamefont {G.}~\bibnamefont {Cao}}, \bibinfo
  {author} {\bibfnamefont {T.}~\bibnamefont {Wang}}, \bibinfo {author}
  {\bibfnamefont {J.}~\bibnamefont {Zhang}}, \bibinfo {author} {\bibfnamefont
  {X.}~\bibnamefont {Hu}},\ and\ \bibinfo {author} {\bibnamefont {et~al.}},\
  }\bibfield  {title} {\bibinfo {title} {Anisotropic g-factor and spin–orbit
  field in a germanium hut wire double quantum dot},\ }\href
  {https://doi.org/10.1021/acs.nanolett.1c00263} {\bibfield  {journal}
  {\bibinfo  {journal} {Nano Letters}\ }\textbf {\bibinfo {volume} {21}},\
  \bibinfo {pages} {3835–3842} (\bibinfo {year} {2021})}\BibitemShut
  {NoStop}%
\bibitem [{\citenamefont {Stepanenko}\ \emph {et~al.}(2012)\citenamefont
  {Stepanenko}, \citenamefont {Rudner}, \citenamefont {Halperin},\ and\
  \citenamefont {Loss}}]{Stepanenko2012}%
  \BibitemOpen
  \bibfield  {author} {\bibinfo {author} {\bibfnamefont {D.}~\bibnamefont
  {Stepanenko}}, \bibinfo {author} {\bibfnamefont {M.}~\bibnamefont {Rudner}},
  \bibinfo {author} {\bibfnamefont {B.~I.}\ \bibnamefont {Halperin}},\ and\
  \bibinfo {author} {\bibfnamefont {D.}~\bibnamefont {Loss}},\ }\bibfield
  {title} {\bibinfo {title} {Singlet-triplet splitting in double quantum dots
  due to spin-orbit and hyperfine interactions},\ }\href
  {https://doi.org/10.1103/PhysRevB.85.075416} {\bibfield  {journal} {\bibinfo
  {journal} {Phys. Rev. B}\ }\textbf {\bibinfo {volume} {85}},\ \bibinfo
  {pages} {075416} (\bibinfo {year} {2012})}\BibitemShut {NoStop}%
\bibitem [{\citenamefont {Nowak}\ \emph {et~al.}(2011)\citenamefont {Nowak},
  \citenamefont {Szafran}, \citenamefont {Peeters}, \citenamefont {Partoens},\
  and\ \citenamefont {Pasek}}]{Nowak2011}%
  \BibitemOpen
  \bibfield  {author} {\bibinfo {author} {\bibfnamefont {M.~P.}\ \bibnamefont
  {Nowak}}, \bibinfo {author} {\bibfnamefont {B.}~\bibnamefont {Szafran}},
  \bibinfo {author} {\bibfnamefont {F.~M.}\ \bibnamefont {Peeters}}, \bibinfo
  {author} {\bibfnamefont {B.}~\bibnamefont {Partoens}},\ and\ \bibinfo
  {author} {\bibfnamefont {W.~J.}\ \bibnamefont {Pasek}},\ }\bibfield  {title}
  {\bibinfo {title} {Tuning of the spin-orbit interaction in a quantum dot by
  an in-plane magnetic field},\ }\href
  {https://doi.org/10.1103/PhysRevB.83.245324} {\bibfield  {journal} {\bibinfo
  {journal} {Phys. Rev. B}\ }\textbf {\bibinfo {volume} {83}},\ \bibinfo
  {pages} {245324} (\bibinfo {year} {2011})}\BibitemShut {NoStop}%
\bibitem [{\citenamefont {Nakaoka}\ \emph {et~al.}(2007)\citenamefont
  {Nakaoka}, \citenamefont {Tarucha},\ and\ \citenamefont
  {Arakawa}}]{Nakaoka2007}%
  \BibitemOpen
  \bibfield  {author} {\bibinfo {author} {\bibfnamefont {T.}~\bibnamefont
  {Nakaoka}}, \bibinfo {author} {\bibfnamefont {S.}~\bibnamefont {Tarucha}},\
  and\ \bibinfo {author} {\bibfnamefont {Y.}~\bibnamefont {Arakawa}},\
  }\bibfield  {title} {\bibinfo {title} {Electrical tuning of the $g$ factor of
  single self-assembled quantum dots},\ }\href
  {https://doi.org/10.1103/PhysRevB.76.041301} {\bibfield  {journal} {\bibinfo
  {journal} {Phys. Rev. B}\ }\textbf {\bibinfo {volume} {76}},\ \bibinfo
  {pages} {041301} (\bibinfo {year} {2007})}\BibitemShut {NoStop}%
\bibitem [{\citenamefont {Prechtel}\ \emph {et~al.}(2015)\citenamefont
  {Prechtel}, \citenamefont {Maier}, \citenamefont {Houel}, \citenamefont
  {Kuhlmann}, \citenamefont {Ludwig}, \citenamefont {Wieck}, \citenamefont
  {Loss},\ and\ \citenamefont {Warburton}}]{Prechtel2015}%
  \BibitemOpen
  \bibfield  {author} {\bibinfo {author} {\bibfnamefont {J.~H.}\ \bibnamefont
  {Prechtel}}, \bibinfo {author} {\bibfnamefont {F.}~\bibnamefont {Maier}},
  \bibinfo {author} {\bibfnamefont {J.}~\bibnamefont {Houel}}, \bibinfo
  {author} {\bibfnamefont {A.~V.}\ \bibnamefont {Kuhlmann}}, \bibinfo {author}
  {\bibfnamefont {A.}~\bibnamefont {Ludwig}}, \bibinfo {author} {\bibfnamefont
  {A.~D.}\ \bibnamefont {Wieck}}, \bibinfo {author} {\bibfnamefont
  {D.}~\bibnamefont {Loss}},\ and\ \bibinfo {author} {\bibfnamefont {R.~J.}\
  \bibnamefont {Warburton}},\ }\bibfield  {title} {\bibinfo {title}
  {Electrically tunable hole $g$ factor of an optically active quantum dot for
  fast spin rotations},\ }\href {https://doi.org/10.1103/PhysRevB.91.165304}
  {\bibfield  {journal} {\bibinfo  {journal} {Phys. Rev. B}\ }\textbf {\bibinfo
  {volume} {91}},\ \bibinfo {pages} {165304} (\bibinfo {year}
  {2015})}\BibitemShut {NoStop}%
\bibitem [{\citenamefont {Voisin}\ \emph {et~al.}(2016)\citenamefont {Voisin},
  \citenamefont {Maurand}, \citenamefont {Barraud}, \citenamefont {Vinet},
  \citenamefont {Jehl}, \citenamefont {Sanquer}, \citenamefont {Renard},\ and\
  \citenamefont {De~Franceschi}}]{Voisin2016}%
  \BibitemOpen
  \bibfield  {author} {\bibinfo {author} {\bibfnamefont {B.}~\bibnamefont
  {Voisin}}, \bibinfo {author} {\bibfnamefont {R.}~\bibnamefont {Maurand}},
  \bibinfo {author} {\bibfnamefont {S.}~\bibnamefont {Barraud}}, \bibinfo
  {author} {\bibfnamefont {M.}~\bibnamefont {Vinet}}, \bibinfo {author}
  {\bibfnamefont {X.}~\bibnamefont {Jehl}}, \bibinfo {author} {\bibfnamefont
  {M.}~\bibnamefont {Sanquer}}, \bibinfo {author} {\bibfnamefont
  {J.}~\bibnamefont {Renard}},\ and\ \bibinfo {author} {\bibfnamefont
  {S.}~\bibnamefont {De~Franceschi}},\ }\bibfield  {title} {\bibinfo {title}
  {Electrical control of g-factor in a few-hole silicon nanowire mosfet},\
  }\href {https://doi.org/10.1021/acs.nanolett.5b02920} {\bibfield  {journal}
  {\bibinfo  {journal} {Nano Letters}\ }\textbf {\bibinfo {volume} {16}},\
  \bibinfo {pages} {88} (\bibinfo {year} {2016})}\BibitemShut {NoStop}%
\bibitem [{\citenamefont {Burkard}\ and\ \citenamefont
  {Imamoglu}(2006)}]{Burkard2006}%
  \BibitemOpen
  \bibfield  {author} {\bibinfo {author} {\bibfnamefont {G.}~\bibnamefont
  {Burkard}}\ and\ \bibinfo {author} {\bibfnamefont {A.}~\bibnamefont
  {Imamoglu}},\ }\bibfield  {title} {\bibinfo {title} {Ultra-long-distance
  interaction between spin qubits},\ }\href
  {https://doi.org/10.1103/PhysRevB.74.041307} {\bibfield  {journal} {\bibinfo
  {journal} {Phys. Rev. B}\ }\textbf {\bibinfo {volume} {74}},\ \bibinfo
  {pages} {041307} (\bibinfo {year} {2006})}\BibitemShut {NoStop}%
\bibitem [{\citenamefont {Jin}\ \emph {et~al.}(2012)\citenamefont {Jin},
  \citenamefont {Marthaler}, \citenamefont {Shnirman},\ and\ \citenamefont
  {Sch\"on}}]{Jin2012}%
  \BibitemOpen
  \bibfield  {author} {\bibinfo {author} {\bibfnamefont {P.-Q.}\ \bibnamefont
  {Jin}}, \bibinfo {author} {\bibfnamefont {M.}~\bibnamefont {Marthaler}},
  \bibinfo {author} {\bibfnamefont {A.}~\bibnamefont {Shnirman}},\ and\
  \bibinfo {author} {\bibfnamefont {G.}~\bibnamefont {Sch\"on}},\ }\bibfield
  {title} {\bibinfo {title} {Strong coupling of spin qubits to a transmission
  line resonator},\ }\href {https://doi.org/10.1103/PhysRevLett.108.190506}
  {\bibfield  {journal} {\bibinfo  {journal} {Phys. Rev. Lett.}\ }\textbf
  {\bibinfo {volume} {108}},\ \bibinfo {pages} {190506} (\bibinfo {year}
  {2012})}\BibitemShut {NoStop}%
\end{thebibliography}%
\end{document}